\documentclass[aps,prd,nofootinbib,showpacs,twocolumn]{revtex4-1}

\usepackage{amstext,amssymb,amsmath}
\usepackage[english]{babel}
\usepackage[ansinew]{inputenc}
\usepackage{graphicx}
\usepackage{units}
\usepackage{hyperref}

\hypersetup{
   pdftitle={Exotic Charges, Multicomponent Dark Matter and Light Sterile Neutrinos},
   pdfauthor={Julian Heeck, He Zhang}
}

\begin{document}

\let \Lold \L
\def \L {\mathcal{L}} 
\let \epsilonold \epsilon
\def \epsilon {\varepsilon} 
\newcommand{\del}{\partial} 
\newcommand{\dd}{\mathrm{d}} 
\newcommand{\matrixx}[1]{\begin{pmatrix} #1 \end{pmatrix}} 
\newcommand{\hc}{\mathrm{h.c.}}
\newcommand{\re}{\mathrm{Re}\,}
\newcommand{\im}{\mathrm{Im}\,}
\newcommand{\diag}{\mathrm{diag}}

\renewcommand{\thefootnote}{\fnsymbol{footnote}} 

\allowdisplaybreaks

\title{Exotic Charges, Multicomponent Dark Matter and Light Sterile Neutrinos}

\author{Julian Heeck}
\email[Electronic address: ]{julian.heeck@mpi-hd.mpg.de}
\affiliation{Max-Planck-Institut f\"ur Kernphysik, Saupfercheckweg 1, 69117 Heidelberg, Germany}

\author{He Zhang}
\email[Electronic address: ]{he.zhang@mpi-hd.mpg.de}
\affiliation{Max-Planck-Institut f\"ur Kernphysik, Saupfercheckweg 1, 69117 Heidelberg, Germany}



\begin{abstract}

Generating small sterile neutrino masses via the same seesaw mechanism that suppresses active neutrino masses requires a specific structure in the neutral fermion mass matrix. We present a model where this structure is enforced by a new $U(1)'$ gauge symmetry, spontaneously broken at the TeV scale. The additional fermions necessary for anomaly cancellation need to carry exotic charges in order not to spoil the neutrino structure and turn out to form multicomponent cold dark matter. The active--sterile mixing then connects the new particles and the Standard Model---opening a new portal in addition to the usual Higgs- and kinetic-mixing portals---which leads to dark matter annihilation almost exclusively into neutrinos.

\end{abstract}

\maketitle


\section{Introduction}
\label{sec:intro}
\addtocounter{footnote}{2}

Neutrino oscillation experiments have by now firmly established the
existence of neutrino oscillations and lepton flavor mixing,
indicating that the Standard Model (SM) framework in particle
physics has to be extended to include neutrino masses. Apart from
the traditional three neutrino oscillation picture, the LSND~\cite{Aguilar:2001ty} and
MiniBooNE~\cite{AguilarArevalo:2010wv} short-baseline experiments suggest the presence of sterile
neutrinos at the eV scale, which do not participate in the weak
interaction but mix with active neutrinos with a mixing angle
$\theta_s \sim \mathcal{O} (0.1)$~\cite{Peres:2000ic,*Sorel:2003hf,*Giunti:2011gz,Kopp:2011qd}.
Furthermore, recent re-evaluations of reactor anti-neutrino fluxes
indicate that the previous reactor neutrino experiments had
observed a flux deficit, which can in fact be interpreted by
additional sterile neutrinos with masses at the eV scale (see
Ref.~\cite{Abazajian:2012ys} for an exhaustive overview of the
field). Moreover, the light element abundances from precision
cosmology and Big Bang nucleosynthesis favor extra radiation in the
Universe, which could be interpreted with the help of one additional
sterile neutrino, albeit with a mass below eV~\cite{Hamann:2010bk,*Izotov:2010ca,*Aver:2010wq,*Hamann:2011ge}.

From the theoretical point of view, the question then arises how the
small sterile neutrino mass scale $\mathcal{O}(\unit{eV})$ can be
motivated compared to the electroweak scale of $\mathcal{O}(\unit[100]{GeV})$. Recall that the seesaw
mechanism~\cite{Minkowski:1977sc,*Yanagida:1979as,*Mohapatra:1979ia,*GellMann:1980vs}
is one of the most popular theoretical attempts to understand
the smallness of active neutrino masses and also the baryon number asymmetry of
the Universe. An obvious ansatz is to use the same seesaw mechanism
to suppress sterile neutrino masses, i.e., to put the sterile neutrinos on the same side of the seesaw as the active neutrinos. To this end, the right-handed
neutrino content has to be extended compared to that in the simplest
type-I seesaw mechanism, and a specific flavor structure, i.e., the
minimal extended seesaw (MES), has to be
employed in order to let the sterile neutrino mass couplings mimic the ones of the active neutrinos~\cite{Barry:2011wb,Zhang:2011vh}. Explicitly, in the MES
model, the SM fermion content is extended by adding three
right-handed neutrinos $\nu_{Ri}$ (for $i=1,2,3$) together with one
singlet fermion $S$,\footnote{To clear up potential confusion right away: all
the $\nu_{R,j}$ and $S_j$ introduced in this paper are just
right-handed fermions, sometimes referred to as singlets. We denote
the $S_j$ with a different symbol than $\nu_{R i}$ to emphasize that they are not the
usual right-handed neutrinos from the seesaw mechanism, because they
carry additional (hidden) quantum numbers and do therefore not partner up with the active neutrinos.} while the full Majorana mass matrix for the neutral fermions in the basis
$(\nu_e,\nu_\mu,\nu_\tau,\nu^c_{R,1},\nu^c_{R,2},\nu^c_{R,3},S^c)$
is given by
\begin{align}
 \mathcal{M}_\mathrm{MES} = \matrixx{0 & m_D & 0 \\ m^T_D & M_R & m_S \\ 0 & m^T_S & 0} .
\label{eq:mes}
\end{align}
After integrating out the heavy right-handed neutrinos $\nu_{R,i}$,
one obtains three massive eigenstates with masses around $M_R$,
three with masses around the eV scale, and one massless neutrino.
Taking a hierarchical spectrum $m_D< m_S \ll M_{R}$, the heaviest of
the three light eigenstates will be a sterile neutrino and its
admixture with active neutrinos is suppressed by a factor
$\mathcal{O}(m_D/m_S)$.\footnote{In the MES framework, the $\nu_e
\leftrightarrow S$ conversion has also been used to solve the solar
neutrino anomaly~\cite{Ma:1995gf,*Chun:1995js}.} See Ref.~\cite{Zhang:2011vh}
and Sec.~\ref{sec:neutrino_masses} for detailed discussions.

Let us briefly comment on a generalization of the MES structure,
promoting $m_D$, $m_S$ and $M_R$ to matrices of dimension
$n(\nu_L)\times n(\nu_R)$, $n(\nu_R)\times n (S)$ and
$n(\nu_R)\times n(\nu_R)$, respectively. In other words, we take
$n(\nu_L)$ active, left-handed neutrinos, $n(S)$ will-be sterile
neutrinos and $n(\nu_R)$ heavy right-handed neutrinos. As far as the
mass matrices are concerned, the $n (S)$ fermions $S^c$ behave just
like the SM neutrinos $\nu_L$---per construction---so we can use the
standard argument~\cite{Schechter:1980gr} to determine the number of
massless states as $n(\nu_L) + n(S) - n(\nu_R)$. Global fits using
neutrino oscillations with $n(\nu_L)+n(S)$ light neutrinos are only
sensitive to mass-squared differences, so one light neutrino is
always allowed to be massless. Consequently, we need at least
$2+n(S)$ heavy right-handed neutrinos $\nu_{R,i}$ if we want $n(S)$
light sterile neutrinos---dubbed $3+n(S)$ scheme. The minimal
case---which lends the MES scheme its name---is then $n(S) = 1$ and
$n(\nu_R) = 3$. This case will be discussed in the main part of this
paper, but we will also comment on the extensions described in this
paragraph.

The MES structure defined in Eq.~\eqref{eq:mes}---meaning the zeros
in the upper and lower right corners of $\mathcal{M}_\mathrm{MES}
$---has to be enforced and motivated by some symmetry. It can be
obtained with discrete flavor symmetries under which the
right-handed neutrinos and $S$ carry different
charges~\cite{Zhang:2011vh}. In addition, the MES structure could
also be obtained in models with abelian symmetries. For example, one
may introduce an extra $U(1)'$ symmetry and take the three
right-handed neutrinos $\nu_{R,i}$ to be neutral under $U(1)'$ (all
SM particles are neutral, too). One may then write down a bare
Majorana mass matrix $M_R$ for $\nu_R$, which is unprotected by the
electroweak or $U(1)'$ scale. The right-handed singlet $S$ on the
other hand carries a $U(1)'$ charge $Y'$, and we further introduce
an SM singlet scalar $\phi$ with charge $-Y'$. The gauge invariant
coupling $\overline{S^c} \nu_R \phi$ then generates the $m_S$ matrix
in Eq.~\eqref{eq:mes} after $\phi$ acquires a vacuum expectation
value (VEV), while the Majorana mass for $S$ (i.e., $\overline{S^c}
S$) and a coupling to the active $\nu_L$ are still forbidden by the
$U(1)'$ symmetry at the renormalizable level. Such a simple
realization of MES however suffers from the problem of triangle
anomalies, and can therefore only work as a global symmetry, whose
spontaneous breaking would result in a massless Goldstone boson.
While this might not be disastrous, more interesting phenomenology
arises when the $U(1)'$ is promoted to a local symmetry.
Consequently, one has to extend the model by additional chiral
fermions so as to cancel the arising gauge anomalies.\footnote{ An
alternative would be the implementation of the Green--Schwarz
mechanism~\cite{Green:1984sg} to cancel anomalies. Since
anomaly-cancelling fermions turn out to have far more interesting
effects, we will not discuss this here.} Along these lines, possible
model constructions for sterile neutrinos in the $U(1)'$ framework
have already been discussed in
Refs.~\cite{Babu:2003is,*Babu:2004mj,*Sayre:2005yh}, using an
effective field theory approach.

In this paper we will work in the seesaw framework and discuss
minimal renormalizable and anomaly-free $U(1)'$ models, which are
spontaneously broken by just one additional scalar and reproduce the
MES structure accounting for the $3+1$ or $3+2$ scheme of light
sterile neutrinos. In particular, we will show that the additional
singlet fermions employed for the anomaly cancellation turn out
to be stable---due to remaining $\mathbb{Z}_N$
symmetries~\cite{Batell:2010bp}---implying that they can be viewed
as good candidates for the dark matter (DM) in the Universe. The
remaining parts of this work are organized as follows: In
Sec.~\ref{sec:dark_symmetry} we explain the framework of exotic
charges under a gauged $U(1)'$ and why they are needed to obtain the MES structure. In Sec.~\ref{sec:pheno} we discuss
in some detail the phenomenology of a specific example with one
light sterile neutrino ($3+1$ scheme) and three stable DM
candidates, with a focus on the novel effects inherent in our model.
We briefly discuss other interesting examples of this framework in
Sec.~\ref{sec:other_charge_assignments}, including an extension to
the $3+2$ case. Finally, we summarize our work in Sec.~\ref{sec:conclusion}.

\section{Dark Symmetry}
\label{sec:dark_symmetry}

As already mentioned in the introduction, adding just one extra
right-handed singlet $S$ to the three $\nu_{R}$'s results in
triangle anomalies if only $S$ is charged under the extra $U(1)'$
symmetry. Instead of treating $U(1)'$ as a global symmetry, we gauge
the $U(1)'$ in the rest of the work, and accordingly introduce additional
singlet chiral fermions to cancel the anomalies. As we will see below, these new states need to decouple from the neutrino sector in order not to spoil the MES structure, and
automatically lead to DM candidates without the need for additional stabilizing discrete
symmetries.

For a gauged $U(1)'$ symmetry under which all SM particles are singlets,
there are no mixed triangle anomalies, so anomaly freedom reduces
to the equations
\begin{align}
\sum_{f} Y'(f) =0 && {\rm and} &&
\sum_{f} (Y'(f))^3 =0 \,, \label{eq:cancel}
\end{align}
where $f$ stands for our new right-handed fermions. In order to cancel the
contribution from $S \equiv S_1$, more $U(1)'$ charged chiral
fermions $S_{i\geq 2}$ have to be introduced. The solutions of
Eq.~\eqref{eq:cancel} for $n=2$ are simply given by
$Y'(S_1)=-Y'(S_2)$. In this case, a bare mass term $ m
\overline{S^c_1} S_2$---unconstrained by any symmetry---can be constructed, which spoils the desired MES structure for light sterile neutrinos unless we make $m$ very small. There is no integer solution for $n=3$ according to the famous Fermat theorem, and it can be shown more generally that Eq.~\eqref{eq:cancel} with $n=3$ only has solutions with one $Y'$ being zero, effectively reducing it to the case with $n=2$. In
the case of $n=4$, it is easy to prove that there is no
phenomenologically interesting solution since two of the $S_i$ must have $U(1)'$ charges of opposite sign and equal magnitude, inducing
an unconstrained bare mass term as in the case of $n=2$.

For $n\geq 5$ however, there exist interesting non-trivial
anomaly-free charge assignments---dubbed exotic charges hereafter---for example the set $(10$, $4$, $-9$, $2$,
$-7)$ for $n=5$~\cite{Babu:2003is,*Babu:2004mj,*Sayre:2005yh,Batra:2005rh,Nakayama:2011dj}.
In order to make all new fermions massive at tree level with just
one scalar $\phi$, even more chiral singlets have to be introduced. For the
$3+1$ scheme discussed in the main text, we add seven singlet
fermions $S_i$ to the model (the $3+2$ scheme discussed in Sec.~\ref{sec:threeplustwo} needs six). The charges of all the ten right-handed
fermions are listed in Tab.~\ref{tab:charges}; they are of course by
no means unique, but serve as a simple illustration of this
framework. We further stress that at least three $U(1)'$ singlet
right-handed neutrinos $\nu_{R,i}$ are needed in order to explain
the observed light neutrino mass-squared differences $\Delta
m_{21}^2$, $\Delta m_{31}^2$ and $\Delta m_{41}^2$---as already mentioned in the introduction---resulting in one massless active neutrino. This is however not a hard prediction of the MES scheme; adding a fourth $\nu_R$ (or even more) to the model makes all light neutrinos massive and does not qualitatively change or
complicate the discussion below. Other interesting charge
assignments with similar overall phenomenology are presented in
Sec.~\ref{sec:other_charge_assignments}.

\begin{table}[tp]
\centering
\begin{tabular}{|c|c|c|c|c|c|c|c|c|c|c|c|}
\hline
 & $\nu_{R,1}$ & $\nu_{R,2}$ & $\nu_{R,3}$ & $S_1$ &  $S_2$ &  $S_3$ &  $S_4$ &  $S_5$ &  $S_6$ &  $S_7$ & $\phi$ \\
\hline
$Y'$ & $0$ & $0$ & $0$ & $11$ &  $-5$ &  $-6$ &  $1$ &  $-12$ &  $2$ &  $9$ & $11$\\
\hline
\end{tabular}
\caption{\label{tab:charges} $U(1)'$ charge assignments of the
right-handed fermions and the scalar $\phi$.}
\end{table}

In the scalar sector, we adopt only one SM singlet scalar $\phi$
with $U(1)'$ charge $11$. We can then write down the following
renormalizable couplings relevant for the neutrino masses
\begin{align}\label{eq:Lm}
\begin{split}
-\L_m &= (m_D)_{ij} \overline{\nu_{L,i}} \nu_{R,j}+
\frac{1}{2}(M_R)_{ij} \overline{\nu^c_{R,i}} \nu_{R,j} + w_i
\phi^\dagger \,\overline{S_1^c} \nu_{R,i} \\
&+ y_1 \phi\,
\overline{S_3^c} S_2 + y_2 \phi \,\overline{S_4^c} S_5 + y_3
\phi^\dagger\, \overline{S_6^c} S_7 + \hc ,
\end{split}
\end{align}
where appropriate sums over $i$ and $j$ are understood. The $m_D$ terms stem from electroweak symmetry breaking using
the usual SM Higgs doublet $H$, while $w_i$ and $y_i$ are Yukawa
couplings. Absorbing phases into the $S_j$ we
can take $y_j$ and one of the $w_j$ to be real, while $M_R$ can taken to be real and diagonal as well. Once $\phi$
acquires a VEV, the $m_S$ matrix will be generated as $m_S = w_j
\langle \phi \rangle$, leading to the MES structure in
Eq.~\eqref{eq:mes} (discussed in detail in Sec.~\ref{sec:neutrino_masses}). The other fermions $S_{2\text{--}7}$ decouple
from $\nu_L$, $\nu_R$ and $S_1$, and actually can be paired together
to form three (stable) Dirac fermions $\Psi_{1,2,3}$, to be discussed in Sec.~\ref{sec:dark_matter}.

Let us briefly comment on a theoretical constraint on the model. An
inherent problem in any gauge theory involving abelian factors is
the occurrence of a Landau pole, i.e., a scale at which the gauge
coupling becomes so large that our perturbative calculations break
down. In our model, the one-loop beta function $\beta$ of the $U(1)'$ gauge
coupling $g'$ takes the form
\begin{align}
\begin{split}
\frac{\dd}{\dd \ln \mu} g' &= \beta = \frac{{g'}^3}{16 \pi^2} b \\
&=
\frac{{g'}^3}{16 \pi^2} \left[ \frac{2}{3} \sum_j (Y'(S_j))^2 +
\frac{1}{3} (Y'(\phi))^2\right] ,
\end{split}
\end{align}
so the Landau pole of $g'$ appears around the scale
\begin{align}
\Lambda_L \simeq \Lambda' \exp \left( \frac{8\pi^2}{b \,(g'
(\Lambda'))^2}\right) ,
\end{align}
where $\Lambda'$ characterizes the $U(1)'$ breaking scale.
Inserting the $U(1)'$ charges given in Tab.~\ref{tab:charges} we
find $b = 315$, whereas for the $3+2$ scheme from
Tab.~\ref{tab:charges2} (which will be discussed later on in
Sec.~\ref{sec:threeplustwo}) we have $b=75$.
For $\Lambda' \simeq 1~{\rm TeV}$ and $\Lambda_L \gtrsim M_\mathrm{Pl}
\simeq \unit[10^{19}]{GeV}$, one obtains the constraints
$g'(\Lambda') \lesssim 0.08$ for the $3+1$ case and
$g'(\Lambda')\lesssim 0.17$ for $3+2$ case. Alternatively, if we
take the cutoff scale of the model to be the right-handed neutrino
mass scale, i.e., $\Lambda_L \gtrsim M_{R} \sim \unit[10^{14}]{GeV}$,
these bounds relax to $g'(\Lambda') \lesssim 0.1$ and $g'(\Lambda')
\lesssim 0.2$ for $3+1$ and $3+2$, respectively. These upper bounds
are stricter than the naive perturbativity bound ${g'}^2/4\pi
\lesssim \mathcal{O}(1)/\max (Y')^2$.

\section{Phenomenological Consequences}
\label{sec:pheno}

In this section we discuss the phenomenology of our new particles,
with an emphasis on the novel effects in our framework.

\subsection{Neutrino Masses and Active--Sterile Mixing}
\label{sec:neutrino_masses}

Let us consider the neutrino masses and active--sterile mixing in the
$U(1)'$ model. After the breaking of the gauge group $SU(3)_C \times
SU(2)_L \times U(1)_Y \times U(1)'$ to $SU(3)_C \times
U(1)_\mathrm{EM}$, the full $13 \times 13$ mass matrix for the neutral fermions in the
basis
$\nu=(\nu^{}_{L,1},\nu^{}_{L,2},\nu^{}_{L,3},\nu^c_{R,1},\nu^c_{R,2},\nu^c_{R,3},S^c_1,S^c_2,S^c_3,S^c_4,S^c_5,S^c_6,S^c_7)$
reads
\begin{align}
\mathcal{M} =\begin{pmatrix}  \left( \mathcal{M}_\mathrm{MES}
\right)_{7\times 7} & 0 \cr 0 & \left(\mathcal{M}_S \right)_{6\times
6}\end{pmatrix}  .
\end{align}
Here, the matrix $\mathcal{M}_\mathrm{MES}$ reproduces the MES
structure from Eq.~\eqref{eq:mes} and
$\mathcal{M}_S$ denotes the mass matrix of $S_{2-7}$, explicitly given as
\begin{align}
\mathcal{M}_S =
\left(
\begin{smallmatrix} 0 & y_1 \langle \phi \rangle & 0 & 0
& 0 & 0  \cr y_1 \langle \phi \rangle & 0 & 0 & 0 & 0 & 0 \cr 0 & 0
& 0 & y_2 \langle \phi \rangle & 0 & 0 \cr 0 & 0 & y_2 \langle \phi
\rangle & 0 & 0 & 0 \cr 0 & 0 & 0 & 0 & 0 & y_3 \langle \phi \rangle
\cr 0 & 0 & 0 & 0 & y_3 \langle \phi \rangle & 0
\end{smallmatrix}
\right) .
\end{align}
Obviously $S_{2\text{--}7}$ decouple from the neutrino sector and can no longer be interpreted as right-handed neutrinos, because they do not mix with the SM neutrinos.
The bare mass term $M_R$ in $\mathcal{M}_\mathrm{MES}$ is unrestricted and can be large, as in the
canonical seesaw case.
We will consider this possibility here by setting $M_R \gg
m_D, m_S$, which leads to the effective low-energy neutrino mass matrix
\begin{align}
 \mathcal{M}_\nu^{4\times 4} \simeq -\matrixx{m_D M_R^{-1} m_D^T & m_D M_R^{-1} m_S \\ m_S^T M_R^{-1} m_D^T & m_S^T M_R^{-1} m_S} ,
\label{eq:lowenergy4times4}
\end{align}
for $(\nu_L, S_1^c)$.\footnote{On a more fundamental level, one can integrate out the heavy right-handed
neutrinos $\nu_R$ at energies $E\ll M_R$ to generate the effective dimension-five Weinberg operators $(m_D)_{i j} (m_D)_{k j}\overline{L}_i \tilde{H} H^\dagger \tilde{L}_k/(\langle H \rangle^2 \,(M_R)_{jj})$, $w_i^2 \, \phi^2 \overline{S}_1 S_1^c/(M_R)_{ii}$ and $(m_D)_{ij}
w_j \, \overline{L}_i \tilde{H} S_1 \phi^\dagger/(\langle H \rangle \,(M_R)_{jj})$,
which were the starting point in
Refs.~\cite{Babu:2003is,*Babu:2004mj,*Sayre:2005yh}.}
Such a mass matrix can be diagonalized by means of a unitary transformation as
$\mathcal{M}_\nu^{4\times 4} = V \,\diag (m_1, m_2, m_3, m_4)
\,V^T$. Phenomenologically, the most interesting situation arises
for $m_S \gg m_D$, since the hierarchical structure of
$\mathcal{M}_\nu^{4\times 4}$ allows us to apply the seesaw
expansion once more, and arrive at the sterile neutrino mass
\begin{align}
m_4 \simeq - m_S^T M_R^{-1} m_S
\end{align}
together with the mass matrix for the three active neutrinos
\begin{align}
\begin{split}
\mathcal{M}_\nu^{3\times 3} &\simeq - m_D M_R^{-1} m_D^T \\
&\quad + m_D
M_R^{-1} m_S\, (m_S^T M_R^{-1} m_S)^{-1} m_S^T M_R^{-1} m_D^T  \\
&= U\,
\diag (m_1, m_2, m_3) \,U^T \,,
\end{split}
\end{align}
diagonalized by $U$.
The $4\times 4$ unitary mixing matrix $V$ is approximately given by
\begin{align}\label{eq:V}
V \simeq \matrixx{ (1- \frac{1}{2} R R^\dagger) U & R \\ -R^\dagger
U & 1 - \frac{1}{2} R^\dagger R }
\end{align}
with the active--sterile mixing vector
\begin{align}
 R = m_D M_R^{-1} m_S \,
(m_S^T M_R^{-1} m_S)^{-1} =  \mathcal{O}(m_D/m_S) \,.
\end{align}
As a rough
numerical estimation, for $m_D \sim \unit[10^2]{GeV}$, $m_S \sim  \unit[5
\times  10^2]{GeV}$ and $M_R \sim  \unit[2\times 10^{14}]{GeV}$,
one obtains the active neutrino mass scale $m_\nu \sim \unit[0.05]{eV}$, the sterile neutrino mass scale $m_s \sim \unit[1.3]{eV}$
together with $R \simeq 0.2$. This is in good agreement with the
current global-fit data, i.e., $|R_1| \simeq 0.15$ and $\Delta
m_{41}^2 \simeq \unit[1.8]{eV}^2$ for the $3+1$
scheme~\cite{Kopp:2011qd}.

For Yukawa couplings of order one, the observed large active--sterile mixing implies the scaling $\langle \phi \rangle/\langle H \rangle \sim 5$--$10$. The new physics scale around TeV is hence not tuned to make LHC phenomenology most interesting, but comes directly from the neutrino sector. Actually---even though we obtain the magic TeV scale---the LHC implications of our model are rather boring, as we only expect small mixing effects in the Higgs and $Z$-boson interactions, to be discussed in the next section.

Let us briefly comment on thermal
leptogenesis~\cite{Fukugita:1986hr} in our framework. In principle,
the additional singlet fermions may spoil the ordinary picture of
leptogenesis since the right-handed neutrinos might predominately
decay to sterile neutrinos instead of active neutrinos. This
drawback can be easily circumvented here by choosing the coupling of
the lightest right-handed neutrino $\nu_{R,1}$ to the new states to
be small, i.e., $w_1 \ll w_{2,3}$. This will not modify the desired
MES structure in the neutrino sector, but sufficiently increase the
branching ratio of $\nu_{R,1}$ into SM particles, so standard
thermal leptogenesis ensues.

\subsection{Bosonic Sector}
\label{sec:potential}

In this subsection we will briefly summarize the behavior of $\phi$
and $Z'$. The scalar potential $W$ in our model takes a simple form
\begin{align}
W= -\mu_H^2 |H|^2 + \lambda_H |H|^4 - \mu_\phi^2 |\phi|^2 +
\lambda_\phi |\phi|^4 + \delta\, |H|^2 |\phi|^2 \, ,
\end{align}
where $H$ denotes the SM Higgs doublet, and $\phi$ can be decomposed
as $\phi = (\re \phi + {\rm i}\, \im \phi)/\sqrt{2}= (\langle\re \phi \rangle + \varphi+ {\rm i}\, \im \phi)/\sqrt{2}$. After symmetry
breaking and in unitary gauge, $\im \phi$ is absorbed by the $Z'$ boson, giving it a mass
$M_{Z'}=|11 g' \langle \phi \rangle|$. Due to the
$\delta$ term in the scalar potential, we have a generic mixing
between the remaining real field $\varphi$ and the neutral scalar $h$
contained in $H$, i.e.,
\begin{align}
\begin{pmatrix}h_1 \cr h_2\end{pmatrix} = \begin{pmatrix} \cos\theta & -\sin\theta \cr \sin\theta & \cos\theta \end{pmatrix} \begin{pmatrix}h \cr \varphi\end{pmatrix} ,
\end{align}
where $h_1$ and $h_2$ are the physical mass eigenstates, and the mixing angle $\theta$
is given by
\begin{align}
\sin2\theta =\frac{\delta  \langle \phi \rangle  \langle H
\rangle}{\sqrt{ \left(\lambda_\phi  \langle \phi \rangle^2
-\lambda_H \langle H \rangle^2 \right)^2 + \left(\delta \langle H
\rangle\langle \phi \rangle\right)^2}} \,.
\end{align}
A non-zero $\delta$---and hence $\theta$---opens the well-known Higgs portal~\cite{Patt:2006fw} for the DM production/annihilation,
which will be discussed in the next section.

The Higgs portal $|\phi|^2 |H|^2$ aside, there is one more
renormalizable, gauge-invariant operator that will induce a coupling
between the SM and DM sectors, namely the kinetic-mixing operator
$\sin\xi \ F^{\mu\nu}_Y
F'_{\mu\nu}$~\cite{Holdom:1985ag,*Holdom:1990xp}. This off-diagonal
kinetic term involving the hypercharge and $U(1)'$ field strength
tensors will induce a coupling of the physical $Z'$ boson to the
hypercharge current. The relevant phenomenology of the resulting
interaction between the SM and DM particles can be found in e.g.~Refs.~\cite{Belanger:2007dx,*Mambrini:2011dw,Chu:2011be}.

\subsection{Dark Matter}
\label{sec:dark_matter}

As we mentioned before, the singlet fermions $S_{2\text{--}7}$ are
the DM candidates in our model. To see this point, we write down the full
Lagrangian for the right-handed singlets $S_2$ and $S_3$:
\begin{align}
\begin{split}
\L_{S_{2,3}} &= {\rm i} \overline{S}_2 \gamma^\mu (\del_\mu - {\rm i}
(-5 g') Z'_\mu ) S_2\\
&\quad + {\rm i} \overline{S}_3 \gamma^\mu (\del_\mu -
{\rm i} (-6 g') Z'_\mu ) S_3 + y_1 (  \phi\, \overline{S}_3^c S_2 +
\hc) \, .
\end{split}
\end{align}
By defining the Dirac field $\Psi_1 \equiv S_2 + S_3^c$, the above
Lagrangian can be rewritten in unitary gauge as
\begin{widetext}
\begin{align}
 \L_{S_{2,3}} = {\rm i} \overline{\Psi}_1 \gamma^\mu \del_\mu \Psi_1 + g' Z'_\mu
\overline{\Psi}_1 \gamma^\mu \left( \frac{(-5)-(-6)}{2} +
\frac{(-5)+ (-6)}{2} \gamma_5 \right) \Psi_1 + \frac{y_1}{\sqrt{2}}
\left(\langle \re (\phi)\rangle + \varphi\right) \ \overline{\Psi}_1 \Psi_1 \, .
\end{align}
After spontaneous symmetry breaking, $\Psi_1$ acquires a Dirac mass
$M_1\equiv -y_1 \langle{\re \phi}\rangle/\sqrt{2}$. Similarly, we
can define $\Psi_2 \equiv S_4 + S_5^c$ and $\Psi_3 \equiv S_6 +
S_7^c$ for $S_{4,5}$ and $S_{6,7}$, and obtain altogether the DM Lagrangian
\begin{align}
\begin{split}
\L_{\mathrm{DM}} &= \sum_{j=1,2,3} \left[ {\rm i} \overline{\Psi}_j \gamma^\mu \del_\mu \Psi_j -M_j   \overline{\Psi}_j \Psi_j - \frac{M_j}{\langle \re (\phi) \rangle} \ \varphi\ \overline{\Psi}_j \Psi_j
 + \frac{ g'}{2} Z'_\mu \overline{\Psi}_j \gamma^\mu \left[ (Y'_{2 j}- Y'_{2 j + 1}) + (Y'_{2 j}+ Y'_{2 j+1})  \gamma_5 \right] \Psi_j \right] ,
\end{split}
\label{eq:DM_lagrangian}
\end{align}
\end{widetext}
where we defined $Y'_j \equiv Y'(S_j)$. The stability of these fields will be discussed below, but let us first take a look at the interactions involving the will-be sterile
neutrino $S_1$, given by the Lagrangian
\begin{align}
\L_{S_1} = {\rm i} \overline{S}_1 \gamma^\mu (\del_\mu - {\rm i} (11
g') Z'_\mu ) S_1 + \left( w_i \phi^\dagger \ \overline{S_1^c}
\nu_{R,i} + \hc\right) .
\end{align}
The important part is the $Z'$ interaction, as it allows for the
annihilation $\Psi_{i} \Psi_{i} \rightarrow Z' \rightarrow S_1 S_1$.
Since the physical sterile neutrino $\nu_s \equiv \nu_4$ consists
mainly of $S_1$, but contains a not-too-small part of the active
neutrinos $\nu_{e,\mu,\tau}$, this process connects the DM to the SM
sector. Specifically, this ``neutrino portal'' takes
the form
\begin{align}
\begin{split}
 \L_{\nu \text{-}\mathrm{portal}} &= \frac{g'}{2} Z'_\mu \Bigg[\overline{\Psi}_1 \gamma^\mu (1 - 11 \gamma_5 ) \Psi_1\\ & \quad+\overline{\Psi}_2 \gamma^\mu (13 - 11 \gamma_5 ) \Psi_2 + \overline{\Psi}_3 \gamma^\mu (-7 + 11 \gamma_5 ) \Psi_3 \\
& \quad+ 11 \sum_{i,j = 1}^4 V_{4 i}^* V_{4 j}  \ \left(\overline{\nu}_i \gamma^\mu \gamma_5 \nu_j+\overline{\nu}_i \gamma^\mu \nu_j\right) \Bigg] ,
\end{split}
\end{align}
where the four light mass eigenstates $\nu_j$ are written as
Majorana spinors and the unitary matrix $V$ is defined in
Eq.~\eqref{eq:V}.

We further note that the heavier DM particles can also convert
to the lighter ones, i.e., $\Psi_i \Psi_i \rightarrow \Psi_j \Psi_j$
via the $s$-channel exchange of $Z'$ or $\phi$. Moreover, $\Psi_i$
may also annihilate to $Z'$ and $\phi$, which can enhance the
total annihilation cross section significantly.

\begin{figure}[t]
\includegraphics[width=.45\textwidth]{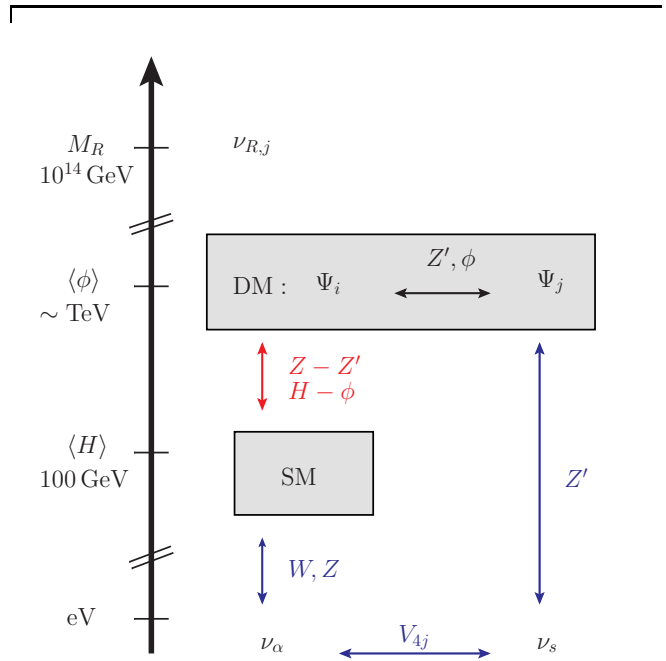}
\caption{\label{fig:1} Visualization of the different scales in our
framework, as well as the relevant interaction channels. The red
connection between the DM and SM sectors represents the well-known
kinetic-mixing and Higgs portals (based on vector and scalar mixing
respectively), while the blue interactions are relevant for the neutrino portal (based on fermion mixing). Interactions with
the $\nu_{R,j}$ are highly suppressed and not shown.}
\end{figure}

The model content and relevant scales are illustrated in
Fig.~\ref{fig:1}. The (self-interacting) DM sector couples to the SM
just like all models with a dark symmetry $U(1)_\mathrm{DM}$, namely through
scalar mixing (Higgs portal, parameterized through $\delta$) and vector mixing (kinetic-mixing
portal, parameterized through $\xi$). However, due to the gauge interactions of the DM with the
sterile neutrinos, a new portal through fermion mixing (neutrino portal) opens up in
our model. Since this portal is not often discussed in the literature (see however Refs.~\cite{Falkowski:2009yz,*Lindner:2010rr,Farzan:2011ck}), we will focus on it in the remainder of this paper.

\subsubsection{Stability}
\label{sec:stability}

It is obvious that the $\Psi_i$ fields in
Eq.~\eqref{eq:DM_lagrangian} are stable, since there exists an
accidental global $U(1)^3$ symmetry shifting the phases of $\Psi_i$.
The occurrence of several stable DM particles (multicomponent DM)
results in numerous interesting effects (see
Refs.~\cite{Boehm:2003ha,*Ma:2006uv,*Hur:2007ur,*Cao:2007fy} for
some early work). The underlying reason for the stability in our
case is the remaining exact $\mathbb{Z}_{11}$ symmetry after the
spontaneous breakdown of the $U(1)'$. The $\Psi_j$ form
representations under this discrete gauge group with charges $6$,
$1$ and $2$ (modulo $11$), which stabilizes at least the lightest of
them, even when higher-dimensional operators are considered.

While our model is renormalizable, we expect it to be only valid up
to a certain cutoff scale $\Lambda$, either because quantum gravity
takes over, or because sooner or later we will hit the $U(1)'$
Landau pole---as discussed in Sec.~\ref{sec:dark_symmetry}. At the
cutoff scale, higher-dimensional operators might be generated, and
in our models these will always include $\phi^2 \overline{S}_1
S_1^c/\Lambda$ and the Weinberg operator~\cite{Weinberg:1979sa} for $\nu_L$-Majorana
masses. Taking $\Lambda\sim M_\mathrm{Pl}$ does not destroy the
discussed MES structure if $\langle \phi \rangle \lesssim
\unit[10]{TeV}$. For the charge assignment here, there are also
dimension-six operators like $\overline{S}^c_2 S_4 \overline{S}_6
S_3^c/\Lambda^2$, which break the global $U(1)^3$ to a $U(1)$
symmetry, so only one stable Dirac fermion survives. However, since
these operators are highly suppressed for $\Lambda\sim
M_\mathrm{Pl}$, the resulting lifetimes are typically longer than
the age of the universe, and thus we will not include them in our
discussions below, but take all three $\Psi_j$ to be independently
stable.

\subsubsection{Relic Density and Thermal History}
\label{sec:thermal_evolution}

We will now discuss the interplay of the three portals (Higgs-,
kinetic-mixing-, and neutrino portal) and identify some valid
regions in the parameter space where the correct relic density for
$\Psi_j$ can be obtained. Note that the mixing parameters $\delta$
and $\xi$ are the only new physics parameters we assume to be small
in this paper, all other couplings are somewhat ``natural''. We
restrict ourselves to small mixing parameters solely for simplicity,
as larger values lead to very constrained effects, see
Refs.~\cite{Belanger:2007dx,*Mambrini:2011dw,Chu:2011be,Patt:2006fw,LopezHonorez:2012kv}.
A more detailed parameter scan (including thermal leptogenesis) will
be performed in a separate publication~\cite{preparation}.

We only consider freeze-out scenarios. Also note that we always end up with a thermalized sterile neutrino at the epoch of
neutrino decoupling, so the usual cosmological bounds on
$N_\mathrm{eff}$ and $\sum m_\nu$ hold~\cite{Abazajian:2012ys}. This is to be expected in
models with light sterile neutrinos, and can be solved on the
astrophysics side---as the bounds strongly depend on the used model and datasets~\cite{Joudaki:2012uk}, and of course the underlying cosmology~\cite{Motohashi:2012wc}---or by choosing a smaller-than-eV mass for the sterile neutrino.

\paragraph*{Case A: $\delta,\xi = 0$.}

\begin{figure*}[t]
\includegraphics[width=.44\textwidth]{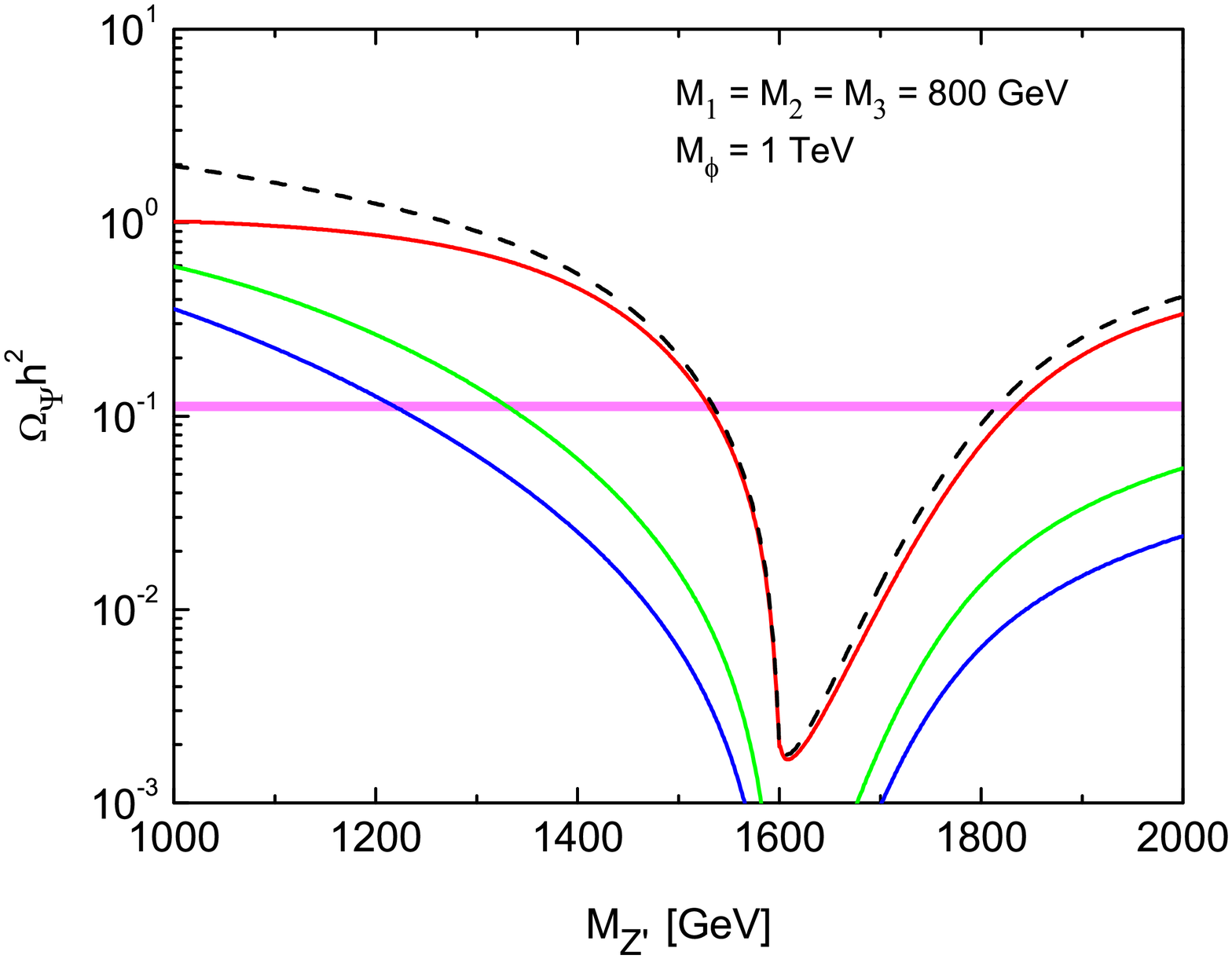}
\includegraphics[width=.44\textwidth]{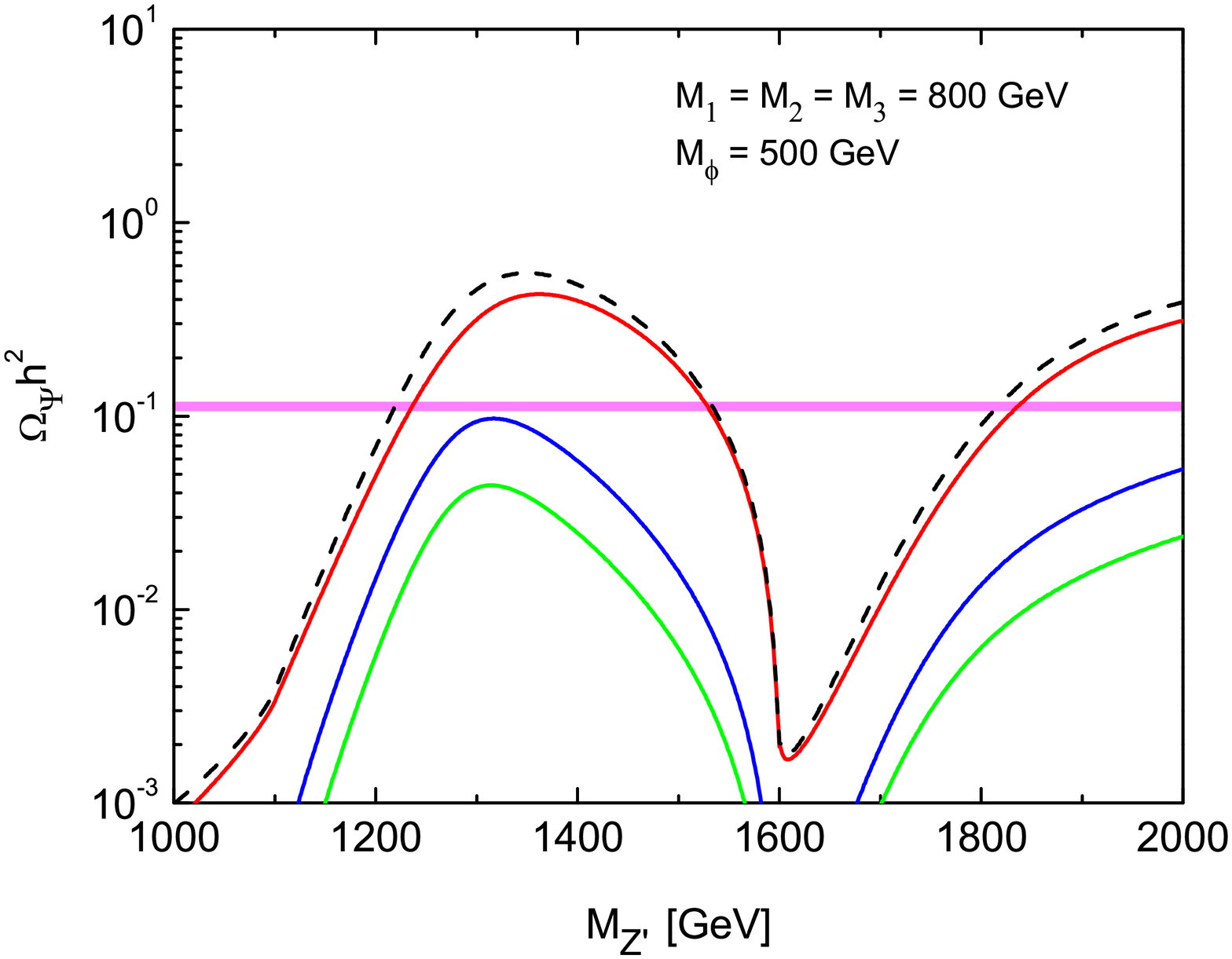}
\includegraphics[width=.44\textwidth]{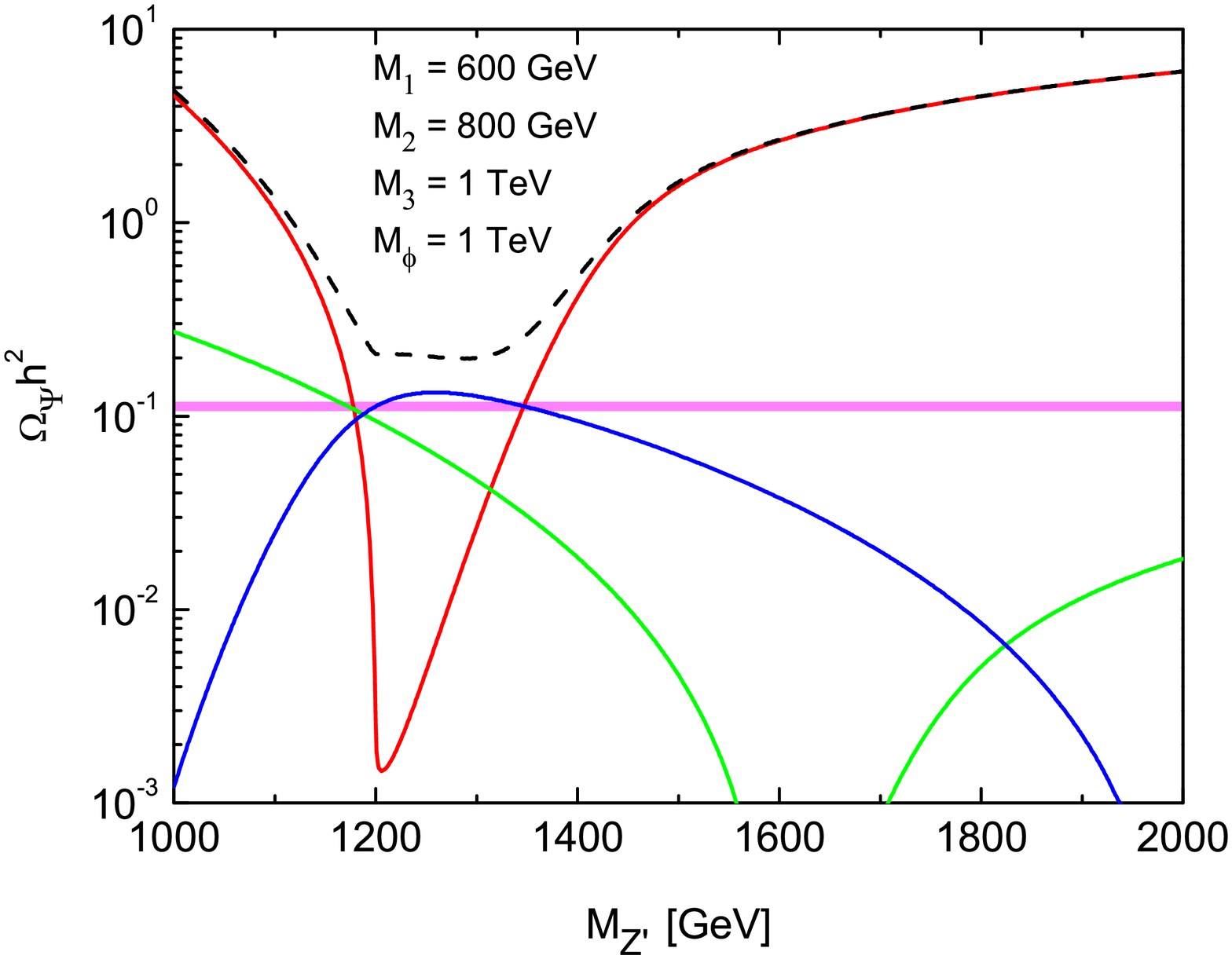}
\includegraphics[width=.44\textwidth]{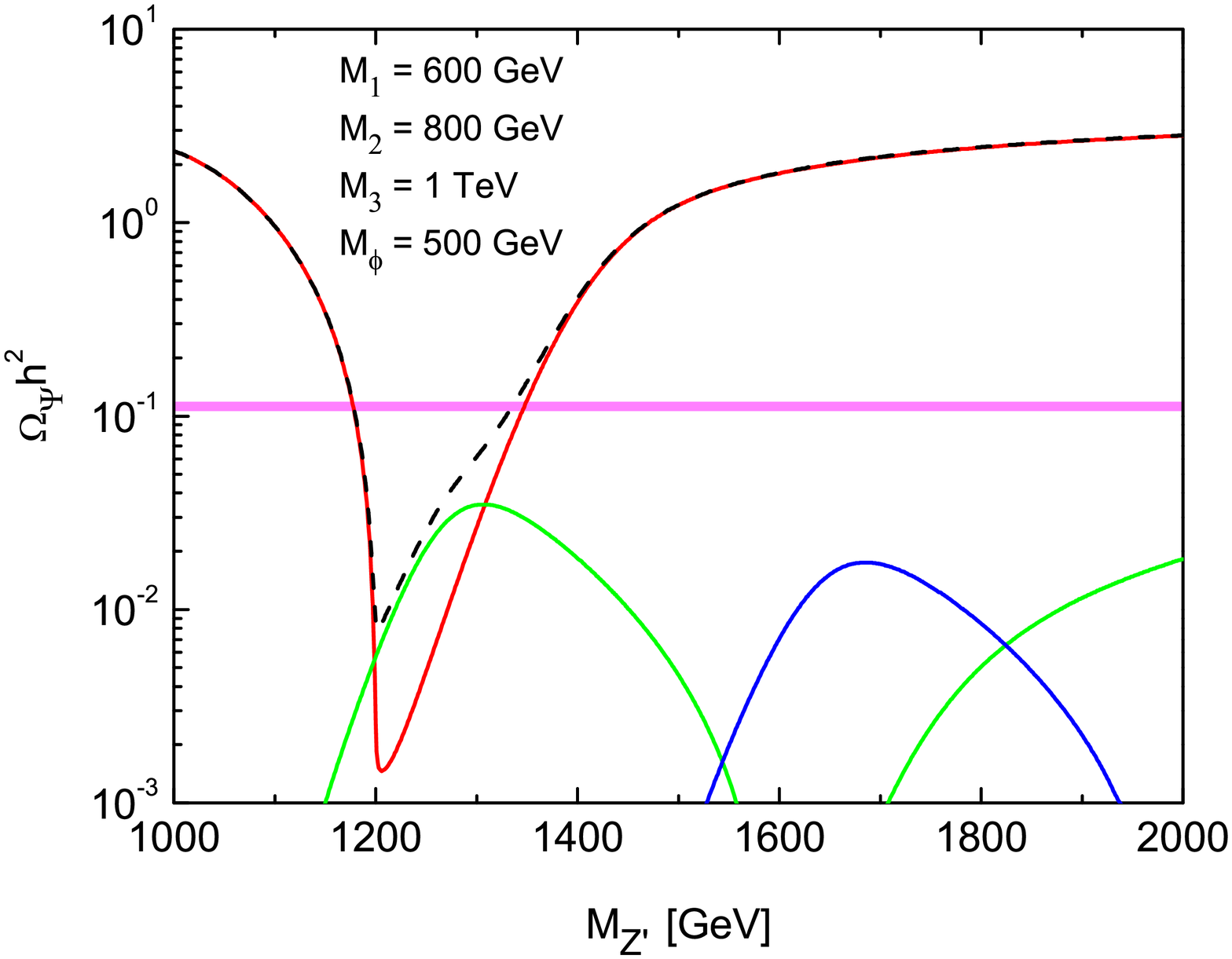}
\includegraphics[width=.44\textwidth]{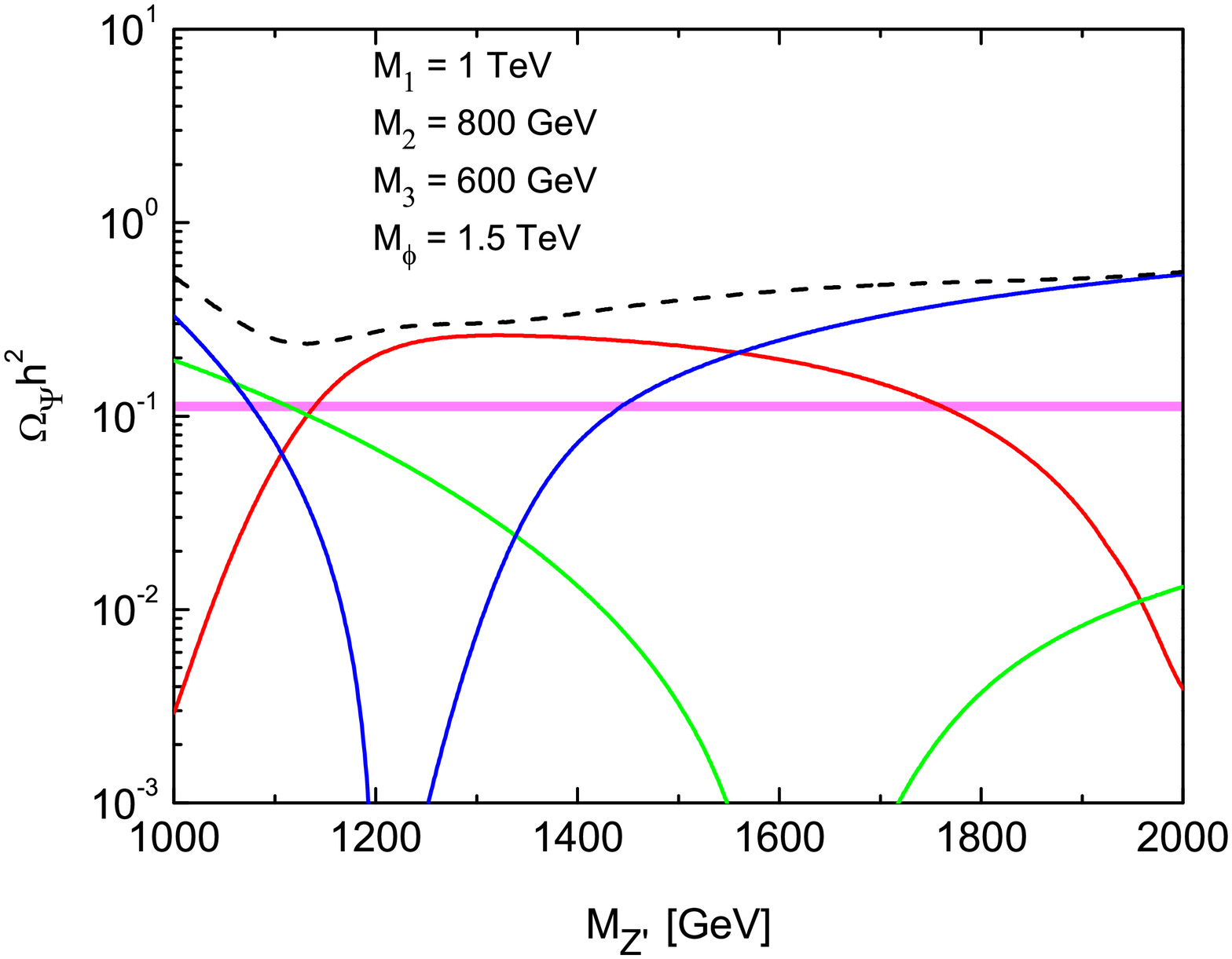}
\includegraphics[width=.44\textwidth]{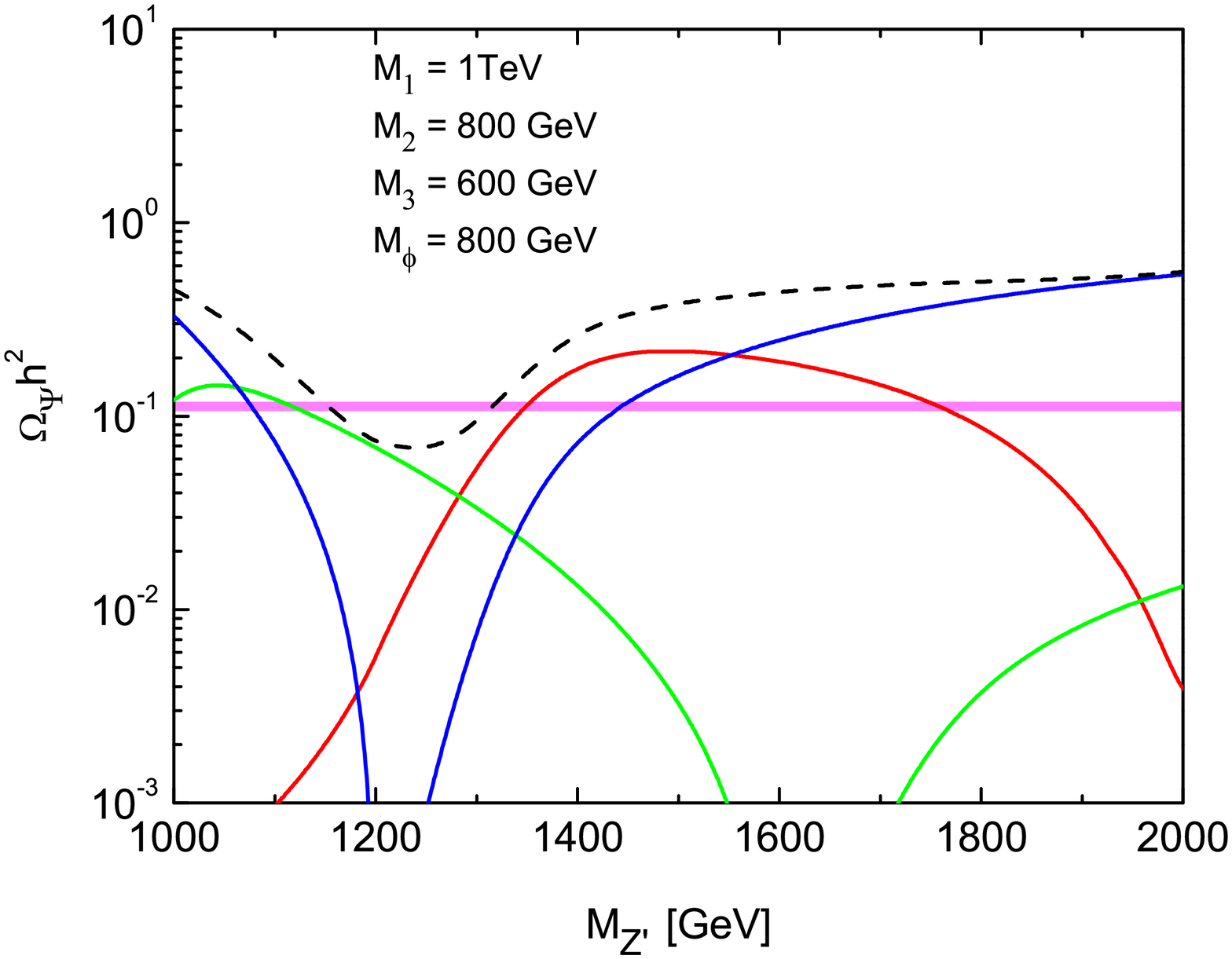}
\caption{\label{fig:2} $\Omega_\Psi h^2$ versus the $Z'$ mass
$M_{Z'}$ for degenerate (top panels) and hierarchical (middle and
bottom panels) DM masses. The VEV was set to $\langle \phi \rangle
=\unit[1.5]{TeV}$ and the scalar mass is labeled on the plot. The
red, green and blue lines show the relic density of $\Psi_1$,
$\Psi_2$ and $\Psi_3$ respectively, while the black dashed line
gives the full $\Omega_\Psi h^2 \equiv \sum_j \Omega_{\Psi_j} h^2$. The horizontal pink band represents the relic density measured by WMAP~\cite{Komatsu:2010fb,*Larson:2010gs} ($1\sigma$ range).}
\end{figure*}

To check the validity of the neutrino portal, we first turn off the
Higgs- and kinetic-mixing portals by setting $\delta = \xi = 0$ (or
at least small enough to be negligible). In this case, the only
connection between the new physics sector and the SM comes from
active--sterile mixing, or, at a more fundamental level, from the
exchange of heavy right-handed neutrinos. Integrating out the
$\nu_R$ gives for example the operator $L H S_1 \phi /M_R$ (using
order one Yukawa couplings), which gives a rough scattering rate for
$ L H \leftrightarrow S_1 \phi$ around $\sim T^3/M_R^2$---to be
compared to the expansion rate in the early universe $\sim
\sqrt{g_*} T^2/M_\mathrm{Pl}$---which puts all particles in equilibrium above
$T \gtrsim \unit[10^{10}]{GeV}$.\footnote{We assume a sufficiently high reheating
temperature after inflation.} Below that temperature,
the two sectors SM and DM (the latter consisting of $Z'$, $\phi$ and $S_j$)
evolve independently, while the temperature decreases due to
expansion of the Universe in both sectors. Nothing really happens
until $T\sim \unit{TeV}$, when the $\Psi_j$ freeze-out occurs.
For simplicity we will ignore the multicomponent structure of the $\Psi_j$ in this qualitative discussion, but will come back to it later on. For now, we assume that the heavier
$\Psi_j$ annihilate sufficiently fast into the lightest $\Psi_j$, which then becomes our DM. This can be accomplished via the mass spectrum of the $\Psi_j$ and $\phi$, see Fig.~\ref{fig:2} for illustrations. To deplete the abundance of the remaining $\Psi_j$ fast enough, we can make use of
the neutrino portal, i.e., the annihilation of the lightest $\Psi_j$
into $\nu_s$ around the $Z'$ resonance.

After freeze-out, we then have overall three decoupled sectors---SM,
$\Psi_j$ and $\nu_s$---all with different temperatures. Above
active-neutrino decoupling, the Universe was radiation dominated,
so only the temperature of $\nu_s$ and the relativistic degrees of
freedom in the SM sector are of interest and will be calculated now.
Using conservation of entropy in the two sectors SM and DM, we have
the equalities
\begin{align}
g_*^\mathrm{SM} T^3 a^3\bigg \vert_{t_\mathrm{sep}} =
g_*^\mathrm{SM} T_\mathrm{SM}^3 a^3\bigg \vert_{t_\mathrm{f}}
\end{align}
and
\begin{align}
g_*^\mathrm{DM} T^3 a^3\bigg \vert_{t_\mathrm{sep}} =
g_*^\mathrm{DM} T_\mathrm{DM}^3 a^3\bigg \vert_{t_\mathrm{f}} ,
\end{align}
where $g_*^X$ denotes the effective number of relativistic degrees
of freedom in sector $X$, $a$ the scale factor, $t_\mathrm{sep}$ the
time when the two sectors just separated from equilibrium (i.e., at
temperatures around $\unit[10^{10}]{GeV}$), and $t_f$ the final time we
are interested in, namely close to active-neutrino decoupling
(e.g.~when $T_\mathrm{SM} \sim \unit[10]{MeV}$). At $t_f$, the SM
sector consists of photons, electrons and neutrinos, while the DM
sector only has the relativistic $S_1 \sim \nu_s$, so we find
\begin{align}
\begin{split}
T_{\nu_s} /T_\mathrm{SM} \bigg \vert_{t_\mathrm{f}} &= \left(
\frac{g_*^\mathrm{DM}(t_\mathrm{sep})}{g_*^\mathrm{DM}(t_f)}
\frac{g_*^\mathrm{SM}(t_f)}{g_*^\mathrm{SM}(t_\mathrm{sep})}\right)^{1/3}\\
&= \left( \frac{65/4}{7/4} \frac{43/4}{427/4}\right)^{1/3} \simeq
0.98 \,.
\end{split}
\end{align}
Ignoring active--sterile oscillations, this would make the sterile
neutrinos slightly colder than the active ones at decoupling, alleviating
cosmological constraints to some degree (the one sterile
neutrino effectively contributes only $\Delta N_\mathrm{eff} = (T_{\nu_s} /T_\mathrm{SM})^4\simeq 0.92$
additional neutrinos to the energy density). However, for the sterile neutrino parameters relevant for the short-baseline anomalies, i.e.~$m_s \sim \unit{eV}$, $\theta_s \sim 0.1$, active--sterile
oscillations will become
effective around $T\sim
\unit[100]{MeV} $--$\unit[1]{MeV}$~\cite{Abazajian:2004aj,*Melchiorri:2008gq,*Giunti:2011cp,*Hannestad:2012ky,*Archidiacono:2012ri},
once again connecting the SM bath and $\nu_s$ and thus thermalizing the
sterile neutrino at neutrino decoupling. Note that the usual
discussions of active--sterile oscillations at these temperatures are
not readily applicable, as our model starts with abundant $\nu_s$
and self-interactions mediated by $Z'$ (freezing out around
$T_\mathrm{DM}\sim \unit[10]{MeV}$). In any case, the cosmological
bound on relativistic degrees of freedom is expected to be
approximately valid in our model and will be discussed in more detail in a separate paper~\cite{preparation}.

\paragraph*{Case B: $\delta \neq 0$.}

Let us open the Higgs portal. The thermal evolution is similar to
case A, but values $\delta \gtrsim 10^{-7}$ will put $\phi$ in
equilibrium with the SM at temperatures below $T\sim
\unit[10]{TeV}$, because the scattering rate $h h \leftrightarrow
\phi \phi$ goes with $\delta^2 T/4 \pi$~\cite{Chu:2011be}. $\phi$
and the rest of the DM sector ($Z'$, $\Psi_j$ and $\nu_s$) are in
equilibrium through $U(1)'$ gauge interactions (for not too small
gauge coupling $g'$), so SM and DM are in equilibrium around DM
freeze-out. For the freeze-out we can again use the neutrino portal,
i.e., resonant annihilation $\Psi \Psi \rightarrow Z'\rightarrow
\nu_s \nu_s$. As $\phi$ and $Z'$ go out of equilibrium around the
same time, the connection between the SM sector and $\nu_s$ is
severed and the two evolve independently for a while, until they are
reconnected around $T\sim \unit[10]{MeV}$ by active--sterile neutrino
oscillations.

In a different region of parameterspace, we can make use of the
resonant annihilation of DM into SM particles via scalars, i.e., the
Higgs portal in the way it is intended. The discussion is then
completely analogous to other $U(1)_\mathrm{DM}$ models, so we refer
the interested reader to Ref.~\cite{LopezHonorez:2012kv} for a
recent evaluation.

\paragraph*{Case C: $\xi \neq 0$.}

A very similar discussion can be made for an open kinetic-mixing
portal. Again small values $\xi \gtrsim 10^{-7}$ suffice to reach
thermal equilibrium of the SM and DM sectors, e.g.~through
scattering $Z h \leftrightarrow Z' h$. The thermal evolution then
closely resembles that of case B, with some minor differences: The
$Z'$--$Z$ mixing couples $\nu_s$ to the SM, so $Z'$ interactions
keep $\nu_s$ thermalized a while longer before it decouples and
finally reconnects with the SM. Furthermore, the DM annihilation
around the $Z'$ resonance contains a small branching ratio into SM
particles.

The above discussion of the cases A, B, and C gives a qualitative overview over the behavior of the sterile neutrino and the DM particles. In all cases, the SM and DM sectors are in equilibrium at some point---creating DM particles, which then freeze out. Even ignoring the Higgs- and kinetic-mixing portals, we can use the neutrino portal to get the correct relic density for $\Psi$. This reheats the sterile neutrinos, but since they invariably re-equilibrate with the active neutrinos---before active-neutrino decoupling---this does not lead to new effects. Knowing that $\Psi$ will have a similar temperature as the SM sector before freeze-out, and that the final-state sterile neutrinos will re-equilibrate with the active neutrinos anyway, the most interesting part left to discuss is then the annihilation $\Psi \Psi \rightarrow \nu_s \nu_s$. For this we again ignore the effects of the Higgs and kinetic-mixing portals for simplicity. We are mainly concerned with the multicomponent aspect of our DM, i.e., whether the right relic density can be obtained for an arbitrary mass spectrum, and which $\Psi$ will be most abundant.

In order to illustrate the feasibility of the DM candidates via the
neutrino portal, we implement the model in
micrOMEGAs~\cite{Belanger:2006is,*Belanger:2008sj,*Belanger:2010pz}
and evaluate the relic density of DM particles $\Psi_i$.
The scalar VEV is
taken to be $\langle \phi \rangle =\unit[1.5]{TeV}$ as an example. The
gauge coupling $g'$ are therefore obtained from the relation of $Z'$
mass and $\langle \phi \rangle$. As shown in the upper panels of
Fig.~\ref{fig:2}, a resonance appears at $M_{\Psi} \simeq M_{Z'}/2
$, and the relic density $\Omega_{\Psi}h^2 \simeq
0.1$ measured by WMAP~\cite{Komatsu:2010fb,*Larson:2010gs} can be achieved. In the
degenerate case (i.e., $M_1 \simeq M_2 \simeq M_3$), the $\Psi_1$
contribution to $\Omega_\Psi h^2$ is dominating because it has the
smallest $Z'$ coupling. Moreover, in case of a small scalar mass,
e.g., $M_\phi =\unit[500]{GeV}$, a new channel $\Psi \Psi \rightarrow
Z' \phi$ is open for light $Z'$, which is observed from the
upper-right panel of Fig.~\ref{fig:2}. For the case of
non-degenerate spectrum (i.e., $M_1 \neq M_2 \neq M_3$), the most
significant contribution to the relic density may come from either
$\Psi_1$, $\Psi_2$ or $\Psi_3$, depending on the specific fermion
spectrum as well as the scalar and vector masses.  As can be seen in the
middle and lower panels of Fig.~\ref{fig:2}, the $\Psi_1$ contribution to the relic density typically dominates, but there exist model parameters that make $\Psi_2 $ or $\Psi_3$ the main DM particle.

\subsubsection{Direct and Indirect Detection}

\begin{figure}[t]
\includegraphics[width=.44\textwidth]{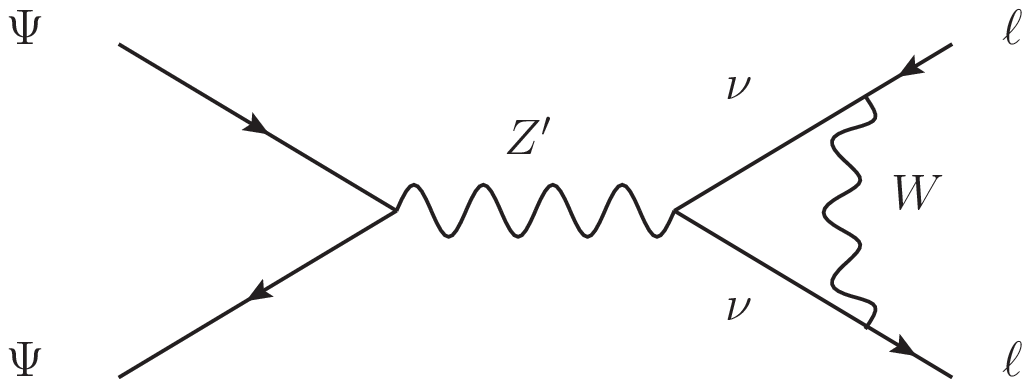}
\caption{\label{fig:3} Coupling of DM to leptons.}
\end{figure}

The neutrino portal discussed so far does not lead to any direct
detection signals, because the cross sections are highly suppressed.
Loop processes connecting $\Psi$ to SM fermions,
e.g.~as in Fig.~\ref{fig:3}, vanish in case of degenerate active--sterile
masses, so these amplitudes are suppressed by tiny factors like
$\Delta m_{41}^2/\mathcal{O}(\unit[100]{GeV})^2 \sim 10^{-22}$.

Indirect detection might naively be more fruitfull, because the
annihilation of the $\Psi_j$ in the Galactic Center or halo leads to two
back-to-back neutrinos with energies $\simeq M_j$ (whichever
$\Psi_j$ is sufficiently abundant), which is an ideal signal for
neutrino telescopes like IceCube.\footnote{The DM--nucleon cross
section in our model is too small to efficiently capture DM inside
the Sun or Earth, so we have to rely on astrophysical objects with
high DM density.} However, since we considered $\Psi_j$ to be a
thermal relic, the self-annihilation cross section is set by the
relic density, which is too small to be probed~\cite{Abbasi:2011eq,*Abbasi:2012ws}---even though the branching ratio into
neutrinos is $\simeq 100\%$, so the signal is as clear as it gets.

Direct and indirect detection measurements are of course sensitive to the Higgs- and kinetic-mixing portal parameters $\delta$ and $\xi$, as discussed in the literature---e.g.~Ref.~\cite{Farzan:2011ck} discusses the Higgs portal in a framework similar to the neutrino portal.

\section{Other Interesting Charge Assignments}
\label{sec:other_charge_assignments}

Having focussed on one specific example using the charges from
Tab.~\ref{tab:charges}, we will now briefly present other charge
assignments with interesting phenomenology. In all cases, we only
introduce one additional scalar $\phi$, so the results concerning
scalar and vector interactions remain unchanged---different
numerical values for the charges aside. Only the sterile neutrino
and dark matter sector will be slightly modified.

\subsection{More Light Sterile Neutrinos}
\label{sec:threeplustwo}

The introduction of $n\geq 2$ light sterile neutrinos ($3+n$ scheme) increases the number of new parameters and most importantly allows for CP-violation in the effective oscillation analysis~\cite{Karagiorgi:2006jf}. This feature can significantly improve the fit to neutrino oscillation data and has been studied extensively~\cite{Peres:2000ic,*Sorel:2003hf,Kopp:2011qd,Donini:2012tt}. Note that the tension with
standard $\Lambda$CDM cosmology typically worsens, depending on the used data sets~\cite{Joudaki:2012uk}.

We can easily modify the above $U(1)'$ framework to accommodate the
$3+2$ MES scheme, by choosing different charges for the ten
right-handed neutrinos: we need at least one more neutral
$\nu_{R,4}$ to generate the necessary light mass squared
differences. Now we have to find charges that treat two of $S_i$ the
same (without loss of generality $S_{1}$ and $S_2$), i.e., $Y'(S_1)
= Y'(S_2)$, so these will become our two light neutrinos after
coupling them to a scalar $\phi$. We can once again find exotic
charges in such a way that the decoupled $S_j$ become massive by
coupling to the same scalar, the magic number for this to happen
seems to be six. See Tab.~\ref{tab:charges2} for a valid
anomaly-free charge assignment with the desired properties---previously
used in Ref.~\cite{Babu:2003is}.
\begin{table}[t]
\centering
\begin{tabular}{|c|c|c|c|c|c|c|c|c|c|c|c|}
\hline
 & $\nu_{R,1}$ & $\nu_{R,2}$ & $\nu_{R,3}$ & $\nu_{R,4}$ & $S_1$ &  $S_2$ &  $S_3$ &  $S_4$ &  $S_5$ &  $S_6$ & $\phi$  \\
\hline
$Y'$ & $0$ & $0$ & $0$ & $0$ &  $-5$ &  $-5$ &  $-1$ &  $6$ &  $2$ &  $3$ & $5$\\
\hline
\end{tabular}
\caption{\label{tab:charges2} Exotic $U(1)'$ charge assignments of the right-handed fermions and the scalar $\phi$ to obtain the $3+2$ MES scheme.}
\end{table}
After breaking the $U(1)'$ and the electroweak symmetry, the
$13\times 13$ mass matrix for the neutral fermions takes the form
\begin{align}
\mathcal{M} =\begin{pmatrix}  \left( \mathcal{M}_\mathrm{MES}
\right)_{9\times 9} & 0 \cr 0 & \left(\mathcal{M}_S \right)_{4\times
4}\end{pmatrix}  ,
\end{align}
where the $9\times 9$ matrix $\mathcal{M}_\mathrm{MES}$ in the basis
$(\nu_e,\nu_\mu,\nu_\tau,\nu^c_{R,1},\nu^c_{R,2},\nu^c_{R,3},\nu^c_{R,4},S_1^c,S_2^c)$
is the obvious extension of the MES structure from
Eq.~\eqref{eq:mes} for the $3+2$ scheme, while $\mathcal{M}_S$
denotes the mass matrix of $S_{3\text{--}6}$, i.e.,
\begin{align}
\mathcal{M}_S =\begin{pmatrix} 0 & y_1 \langle \phi \rangle & 0 & 0
  \cr y_1 \langle \phi \rangle & 0 & 0 & 0  \cr 0 & 0
& 0 & y_2 \langle \phi \rangle  \cr 0 & 0 & y_2 \langle \phi
\rangle & 0
\end{pmatrix} ,
\end{align}
resulting in two Dirac fermions, decoupled from the neutrino sector.

Compared to the $3+1$ scheme discussed so far, the scalar sector is
identical, whereas the dark matter sector is slightly modified
because we have only two stable Dirac fermions---protected by the
remaining discrete gauge group $\mathbb{Z}_5$---instead of three, but two light sterile
neutrinos instead of one. This does not influence the qualitative
behavior significantly.

The expressions from Sec.~\ref{sec:neutrino_masses} for the neutrino
masses go through in the same manner, i.e., we still have
\begin{align}
\begin{split}
\mathcal{M}_\nu^{3\times 3} &\simeq - m_D M_R^{-1} m_D^T \\
&\quad + m_D M_R^{-1} m_S\, (m_S^T M_R^{-1} m_S)^{-1} m_S^T M_R^{-1} m_D^T
\end{split}
\end{align}
and
\begin{align}
 \mathcal{M}_{\nu_s}^{2\times 2} \simeq - m_S^T M_R^{-1} m_S
\end{align}
for the masses, where we assumed $m_D \ll m_S \ll M_R$, and $m_D$,
$m_S$ and $M_R$ are $3\times 4$, $4\times 2$, and $4\times 4$
matrices, respectively. The active--sterile mixing is again $\mathcal{O}(m_D/m_S)$,
so the required values $\mathcal{O}(0.1)$ put the $U(1)'$ breaking
scale naturally in the TeV range.

Let us briefly comment on the thermal evolution of the universe in this model. Seeing as the number of degrees of freedom is smaller (larger) at $t_\mathrm{sep}$ ($t_f$) compared to the $3+1$ scheme of Sec.~\ref{sec:thermal_evolution}, the sterile neutrino bath is colder than the SM bath (prior to neutrino decoupling) by a factor of $\simeq 0.75$. Without active--sterile neutrino oscillations, this would mean that the two sterile neutrinos effectively only contribute $\Delta N_\mathrm{eff} \simeq 0.6$ additional neutrino species to the energy density, alleviating cosmological bounds. It is of course to be expected that active--sterile oscillations before neutrino decoupling generate thermal equilibrium among the neutrinos, giving rise to the usual constraints.

For completeness, we also give an assignment for the $3+3$ case, which
has been fitted to the neutrino anomalies in
Ref.~\cite{Conrad:2012qt}. To make at least five light neutrinos
massive, we need five $\nu_R$. A possible charge assignment for nine
$S_j$ is then $(7,7,7,2,-9,-1,-6,-4,-3)$, with one scalar $\phi \sim
7$. This leads to three light sterile neutrinos and three stable
Dirac DM particles---protected by the remaining discrete gauge group $\mathbb{Z}_7$.

\subsection{Majorana Dark Matter}

Having focused on Dirac DM in the main text for no particular reason, we will now give an example with Majorana DM. For the $3+1$ MES scheme, we take the exotic charges $(6,-3,-3,2,-8,-1,7)$ for the $S_i$ and one scalar with charge $Y'(\phi)=6$. The VEV of $\phi$ breaks $U(1)'\rightarrow \mathbb{Z}_6$, $S_1$ will again become the sterile neutrino, while $S_2$ and $S_3$ share the most general Majorana mass matrix---which we can take to be diagonal without loss of generality---resulting in two Majorana fermions $\Psi_{1,2}$. $(S_4,S_5,S_6,S_7)$ share the mass matrix
\begin{align}
\mathcal{M}_S =\begin{pmatrix} 0 & y_1 \langle \phi \rangle & 0 & 0
  \cr y_1 \langle \phi \rangle & 0 & 0 & 0  \cr 0 & 0
& 0 & y_2 \langle \phi \rangle  \cr 0 & 0 & y_2 \langle \phi
\rangle & 0
\end{pmatrix}  ,
\end{align}
resulting in two Dirac fermions $\Psi_{3,4}$; all $\Psi_j$ are decoupled from the neutrino sector. These particles form representations $\Psi_{1,2} \sim 3 \sim (1,0)$, $\Psi_3 \sim 2 \sim (0,2)$, and $\Psi_4 \sim 1 \sim (1,1)$ under $\mathbb{Z}_6 \cong \mathbb{Z}_2 \times \mathbb{Z}_3$, so depending on the mass spectrum, we can obtain a stable Majorana fermion.

\subsection{Unstable Dark Matter}

The charges for the $S_i$ and $\phi$ discussed so far have been
chosen in such a way that the spontaneous breaking of $U(1)'$ leaves
a nontrivial $\mathbb{Z}_N$ that stabilizes the DM candidates. This is
of course not a generic feature of exotic charges, but just a convenient choice to obtain exactly stable particles. Let us briefly comment
on unstable DM candidates: Taking $(1,-10,9,-7,6,-11,12)$ for the
$3+1$ scheme with a scalar $\phi \sim 1$ gives three Dirac DM
candidates---$\Psi_1 = S_2 + S_3^c$, $\Psi_2 = S_4+S_5^c$, and $\Psi_3 =
S_6+S_7^c$---which are independently stable due to an accidental
global $U(1)^3$ symmetry. However, with this charge assignment,
there is no leftover $\mathbb{Z}_N$ symmetry protecting this
stability. We can study higher-dimensional operators similar to the discussion in Sec.~\ref{sec:stability}. For the charge
assignment here, there are already dimension-five operators
\begin{align}
\phi^2 \overline{S}_3 S_4^c/\Lambda\,, && \phi^2 \overline{S}_3^c
S_6/\Lambda\,, && \phi^2 \overline{S}_2 S_7^c/\Lambda\,,
\end{align}
which break the global $U(1)^3$ to a $U(1)$ symmetry, so only one
stable Dirac fermion survives. Even this stability is not exact, as
there are operators like $\phi^6 \overline{S}^c_5 \nu_R/\Lambda^5$ which break
the global $U(1)$ and lead to DM decay. In this particular example---and for $\Lambda \sim M_\mathrm{Pl}$---the decay would be suppressed enough to still allow for valid DM, but in
principle there are charge assignments with decaying DM, or even no
DM candidate at all.

\subsection{More Less-Exotic Charges}

As was shown in Ref.~\cite{Batra:2005rh}, the $U(1)'$ anomalies from $S_1$ can always be cancelled by a (typically large) number of fermions with basic charges $-2$ and $+1$, instead of the small set of exotic charges used so far. For example, the anomaly of $S_1\sim 4$ can be cancelled with ten copies of $S^{-2}$ and sixteen copies of $S^{+1}$, i.e.~one trades the large charge magnitude of a small number of fermions with the small charge magnitude of a large number of fermions. Since this approach might be seen as less exotic---sacrificing however the small number of particles and parameters employed so far---we will comment on it in our framework. Seeing as the number of fermions $S_j^{-2}$ with charge $-2$ is not equal to the number of fermions $S_j^{+1}$ with charge $+1$~\cite{Batra:2005rh}, it does not suffice to introduce just one more scalar $\phi_2\sim 1$ to make them massive, so we need at least two, e.g.~$\phi_2 \sim 1$ and $\phi_3\sim 2$. For all choices, there will be a coupling of either $S^{+1}$ or $S^{-2}$ to the right-handed neutrinos $\nu_R$, for example $\phi_3 \overline{\nu}_R^c S^{-2}$. Consequently, there is no way of making the anomaly-cancelling fermions massive without modifying the MES structure in Eq.~\eqref{eq:mes}. As our motivation was a consistent realization of this structure, we will not discuss these less-exotic charges any further.

\section{Conclusion}
\label{sec:conclusion}

Generating small sterile neutrino masses via the same seesaw mechanism that suppresses active neutrino masses requires a specific structure in the neutral fermion mass matrix. We showed how this so-called MES structure can be obtained from a new spontaneously broken $U(1)'$ symmetry, under which the ``sterile'' neutrino is charged. Heavily mixed eV-scale steriles hint at a $U(1)'$ breaking scale around TeV. Additional anomaly-cancelling fermions need to carry exotic $U(1)'$ charges in order to not spoil the MES structure, which coincidentally stabilizes one or more of them---all without the need for any discrete symmetries. The main connection between this multicomponent dark matter sector and the Standard Model is the active--sterile mixing (neutrino portal). We discussed how the dark matter annihilation almost exclusively into sterile neutrinos can be used to obtain the measured relic density, and also the interplay with the other two portals (Higgs- and kinetic-mixing portals).

We focussed on a few specific examples, but the presented framework of exotic charges obviously provides a rich playground for model building, depending on the used charges and number of new particles. Worthwhile extensions with $U(1)'$ charged SM fermions, e.g.~$B-L$ type symmetries, can be obtained with slightly more complicated scalar sectors and will be discussed elsewhere. See for example Refs.~\cite{Pospelov:2011ha,*Pospelov:2012gm} for a model with an MES sterile coupled to a gauged baryon number symmetry.

Let us make one last comment: noting that the Standard Model fermion content forms a chiral set (i.e.~has exotic charges) of the gauge group $SU(3)_C \times SU(2)_L \times U(1)_Y$, it is a reasonable assumption that a possible hidden sector also has a chiral structure. The simplest example of a chiral hidden sector is then a $U(1)'$ with exotic charges, as discussed in our work.

\begin{acknowledgments}

The authors would like to thank Jisuke Kubo, Michael Duerr, Pei-Hong
Gu, and especially Thomas Schwetz for useful discussions and
comments. This work was supported by the Max Planck Society through
the Strategic Innovation Fund. J.H.~acknowledges support by the
IMPRS-PTFS.

\end{acknowledgments}

\bibliography{bib}

\begin{thebibliography}{63}%
\makeatletter
\providecommand \@ifxundefined [1]{%
 \@ifx{#1\undefined}
}%
\providecommand \@ifnum [1]{%
 \ifnum #1\expandafter \@firstoftwo
 \else \expandafter \@secondoftwo
 \fi
}%
\providecommand \@ifx [1]{%
 \ifx #1\expandafter \@firstoftwo
 \else \expandafter \@secondoftwo
 \fi
}%
\providecommand \natexlab [1]{#1}%
\providecommand \enquote  [1]{``#1''}%
\providecommand \bibnamefont  [1]{#1}%
\providecommand \bibfnamefont [1]{#1}%
\providecommand \citenamefont [1]{#1}%
\providecommand \href@noop [0]{\@secondoftwo}%
\providecommand \href [0]{\begingroup \@sanitize@url \@href}%
\providecommand \@href[1]{\@@startlink{#1}\@@href}%
\providecommand \@@href[1]{\endgroup#1\@@endlink}%
\providecommand \@sanitize@url [0]{\catcode `\\12\catcode `\$12\catcode
  `\&12\catcode `\#12\catcode `\^12\catcode `\_12\catcode `\%12\relax}%
\providecommand \@@startlink[1]{}%
\providecommand \@@endlink[0]{}%
\providecommand \url  [0]{\begingroup\@sanitize@url \@url }%
\providecommand \@url [1]{\endgroup\@href {#1}{\urlprefix }}%
\providecommand \urlprefix  [0]{URL }%
\providecommand \Eprint [0]{\href }%
\providecommand \doibase [0]{http://dx.doi.org/}%
\providecommand \selectlanguage [0]{\@gobble}%
\providecommand \bibinfo  [0]{\@secondoftwo}%
\providecommand \bibfield  [0]{\@secondoftwo}%
\providecommand \translation [1]{[#1]}%
\providecommand \BibitemOpen [0]{}%
\providecommand \bibitemStop [0]{}%
\providecommand \bibitemNoStop [0]{.\EOS\space}%
\providecommand \EOS [0]{\spacefactor3000\relax}%
\providecommand \BibitemShut  [1]{\csname bibitem#1\endcsname}%
\let\auto@bib@innerbib\@empty
\bibitem [{\citenamefont {Aguilar-Arevalo}\ \emph {et~al.}(2001)\citenamefont
  {Aguilar-Arevalo} \emph {et~al.}}]{Aguilar:2001ty}%
  \BibitemOpen
  \bibfield  {author} {\bibinfo {author} {\bibfnamefont {A.}~\bibnamefont
  {Aguilar-Arevalo}} \emph {et~al.} (\bibinfo {collaboration} {LSND
  Collaboration}),\ }\bibfield  {journal} {\bibinfo  {journal} {Phys.Rev.}\
  }\textbf {\bibinfo {volume} {D64}},\ \bibinfo {pages} {112007} (\bibinfo
  {year} {2001}),\ \Eprint {http://arxiv.org/abs/hep-ex/0104049}
  {hep-ex/0104049}\BibitemShut {NoStop}%
\bibitem [{\citenamefont {Aguilar-Arevalo}\ \emph {et~al.}(2010)\citenamefont
  {Aguilar-Arevalo} \emph {et~al.}}]{AguilarArevalo:2010wv}%
  \BibitemOpen
  \bibfield  {author} {\bibinfo {author} {\bibfnamefont {A.}~\bibnamefont
  {Aguilar-Arevalo}} \emph {et~al.} (\bibinfo {collaboration} {MiniBooNE
  Collaboration}),\ }\bibfield  {journal} {\bibinfo  {journal}
  {Phys.Rev.Lett.}\ }\textbf {\bibinfo {volume} {105}},\ \bibinfo {pages}
  {181801} (\bibinfo {year} {2010}),\ \Eprint {http://arxiv.org/abs/1007.1150}
  {1007.1150}\BibitemShut {NoStop}%
\bibitem [{\citenamefont {Peres}\ and\ \citenamefont
  {Smirnov}(2001)}]{Peres:2000ic}%
  \BibitemOpen
  \bibfield  {author} {\bibinfo {author} {\bibfnamefont {O.}~\bibnamefont
  {Peres}}\ and\ \bibinfo {author} {\bibfnamefont {A.~Y.}\ \bibnamefont
  {Smirnov}},\ }\bibfield  {journal} {\bibinfo  {journal} {Nucl.Phys.}\
  }\textbf {\bibinfo {volume} {B599}},\ \bibinfo {pages} {3} (\bibinfo {year}
  {2001}),\ \Eprint {http://arxiv.org/abs/hep-ph/0011054}
  {hep-ph/0011054}\BibitemShut {NoStop}%
\bibitem [{\citenamefont {Sorel}\ \emph {et~al.}(2004)\citenamefont {Sorel},
  \citenamefont {Conrad},\ and\ \citenamefont {Shaevitz}}]{Sorel:2003hf}%
  \BibitemOpen
  \bibfield  {author} {\bibinfo {author} {\bibfnamefont {M.}~\bibnamefont
  {Sorel}}, \bibinfo {author} {\bibfnamefont {J.~M.}\ \bibnamefont {Conrad}}, \
  and\ \bibinfo {author} {\bibfnamefont {M.}~\bibnamefont {Shaevitz}},\
  }\bibfield  {journal} {\bibinfo  {journal} {Phys.Rev.}\ }\textbf {\bibinfo
  {volume} {D70}},\ \bibinfo {pages} {073004} (\bibinfo {year} {2004}),\
  \Eprint {http://arxiv.org/abs/hep-ph/0305255} {hep-ph/0305255}\BibitemShut
  {NoStop}%
\bibitem [{\citenamefont {Giunti}\ and\ \citenamefont
  {Laveder}(2011{\natexlab{a}})}]{Giunti:2011gz}%
  \BibitemOpen
  \bibfield  {author} {\bibinfo {author} {\bibfnamefont {C.}~\bibnamefont
  {Giunti}}\ and\ \bibinfo {author} {\bibfnamefont {M.}~\bibnamefont
  {Laveder}},\ }\bibfield  {journal} {\bibinfo  {journal} {Phys.Rev.}\ }\textbf
  {\bibinfo {volume} {D84}},\ \bibinfo {pages} {073008} (\bibinfo {year}
  {2011}{\natexlab{a}}),\ \Eprint {http://arxiv.org/abs/1107.1452}
  {1107.1452}\BibitemShut {NoStop}%
\bibitem [{\citenamefont {Kopp}\ \emph {et~al.}(2011)\citenamefont {Kopp},
  \citenamefont {Maltoni},\ and\ \citenamefont {Schwetz}}]{Kopp:2011qd}%
  \BibitemOpen
  \bibfield  {author} {\bibinfo {author} {\bibfnamefont {J.}~\bibnamefont
  {Kopp}}, \bibinfo {author} {\bibfnamefont {M.}~\bibnamefont {Maltoni}}, \
  and\ \bibinfo {author} {\bibfnamefont {T.}~\bibnamefont {Schwetz}},\
  }\bibfield  {journal} {\bibinfo  {journal} {Phys.Rev.Lett.}\ }\textbf
  {\bibinfo {volume} {107}},\ \bibinfo {pages} {091801} (\bibinfo {year}
  {2011}),\ \Eprint {http://arxiv.org/abs/1103.4570} {1103.4570}\BibitemShut
  {NoStop}%
\bibitem [{\citenamefont {Abazajian}\ \emph {et~al.}(2012)\citenamefont
  {Abazajian}, \citenamefont {Acero}, \citenamefont {Agarwalla}, \citenamefont
  {Aguilar-Arevalo}, \citenamefont {Albright} \emph
  {et~al.}}]{Abazajian:2012ys}%
  \BibitemOpen
  \bibfield  {author} {\bibinfo {author} {\bibfnamefont {K.}~\bibnamefont
  {Abazajian}}, \bibinfo {author} {\bibfnamefont {M.}~\bibnamefont {Acero}},
  \bibinfo {author} {\bibfnamefont {S.}~\bibnamefont {Agarwalla}}, \bibinfo
  {author} {\bibfnamefont {A.}~\bibnamefont {Aguilar-Arevalo}}, \bibinfo
  {author} {\bibfnamefont {C.}~\bibnamefont {Albright}},  \emph {et~al.}\ }
  (\bibinfo {year} {2012}),\ \Eprint {http://arxiv.org/abs/1204.5379}
  {1204.5379}\BibitemShut {NoStop}%
\bibitem [{\citenamefont {Hamann}\ \emph {et~al.}(2010)\citenamefont {Hamann},
  \citenamefont {Hannestad}, \citenamefont {Raffelt}, \citenamefont
  {Tamborra},\ and\ \citenamefont {Wong}}]{Hamann:2010bk}%
  \BibitemOpen
  \bibfield  {author} {\bibinfo {author} {\bibfnamefont {J.}~\bibnamefont
  {Hamann}}, \bibinfo {author} {\bibfnamefont {S.}~\bibnamefont {Hannestad}},
  \bibinfo {author} {\bibfnamefont {G.~G.}\ \bibnamefont {Raffelt}}, \bibinfo
  {author} {\bibfnamefont {I.}~\bibnamefont {Tamborra}}, \ and\ \bibinfo
  {author} {\bibfnamefont {Y.~Y.}\ \bibnamefont {Wong}},\ }\bibfield  {journal}
  {\bibinfo  {journal} {Phys.Rev.Lett.}\ }\textbf {\bibinfo {volume} {105}},\
  \bibinfo {pages} {181301} (\bibinfo {year} {2010}),\ \Eprint
  {http://arxiv.org/abs/1006.5276} {1006.5276}\BibitemShut {NoStop}%
\bibitem [{\citenamefont {Izotov}\ and\ \citenamefont
  {Thuan}(2010)}]{Izotov:2010ca}%
  \BibitemOpen
  \bibfield  {author} {\bibinfo {author} {\bibfnamefont {Y.}~\bibnamefont
  {Izotov}}\ and\ \bibinfo {author} {\bibfnamefont {T.}~\bibnamefont {Thuan}},\
  }\bibfield  {journal} {\bibinfo  {journal} {Astrophys.J.}\ }\textbf {\bibinfo
  {volume} {710}},\ \bibinfo {pages} {L67} (\bibinfo {year} {2010}),\ \Eprint
  {http://arxiv.org/abs/1001.4440} {1001.4440}\BibitemShut {NoStop}%
\bibitem [{\citenamefont {Aver}\ \emph {et~al.}(2010)\citenamefont {Aver},
  \citenamefont {Olive},\ and\ \citenamefont {Skillman}}]{Aver:2010wq}%
  \BibitemOpen
  \bibfield  {author} {\bibinfo {author} {\bibfnamefont {E.}~\bibnamefont
  {Aver}}, \bibinfo {author} {\bibfnamefont {K.~A.}\ \bibnamefont {Olive}}, \
  and\ \bibinfo {author} {\bibfnamefont {E.~D.}\ \bibnamefont {Skillman}},\
  }\bibfield  {journal} {\bibinfo  {journal} {JCAP}\ }\textbf {\bibinfo
  {volume} {1005}},\ \bibinfo {pages} {003} (\bibinfo {year} {2010}),\ \Eprint
  {http://arxiv.org/abs/1001.5218} {1001.5218}\BibitemShut {NoStop}%
\bibitem [{\citenamefont {Hamann}\ \emph {et~al.}(2011)\citenamefont {Hamann},
  \citenamefont {Hannestad}, \citenamefont {Raffelt},\ and\ \citenamefont
  {Wong}}]{Hamann:2011ge}%
  \BibitemOpen
  \bibfield  {author} {\bibinfo {author} {\bibfnamefont {J.}~\bibnamefont
  {Hamann}}, \bibinfo {author} {\bibfnamefont {S.}~\bibnamefont {Hannestad}},
  \bibinfo {author} {\bibfnamefont {G.~G.}\ \bibnamefont {Raffelt}}, \ and\
  \bibinfo {author} {\bibfnamefont {Y.~Y.}\ \bibnamefont {Wong}},\ }\bibfield
  {journal} {\bibinfo  {journal} {JCAP}\ }\textbf {\bibinfo {volume} {1109}},\
  \bibinfo {pages} {034} (\bibinfo {year} {2011}),\ \Eprint
  {http://arxiv.org/abs/1108.4136} {1108.4136}\BibitemShut {NoStop}%
\bibitem [{\citenamefont {Minkowski}(1977)}]{Minkowski:1977sc}%
  \BibitemOpen
  \bibfield  {author} {\bibinfo {author} {\bibfnamefont {P.}~\bibnamefont
  {Minkowski}},\ }\bibfield  {journal} {\bibinfo  {journal} {Phys.Lett.}\
  }\textbf {\bibinfo {volume} {B67}},\ \bibinfo {pages} {421} (\bibinfo {year}
  {1977})\BibitemShut {NoStop}%
\bibitem [{\citenamefont {Yanagida}(1979)}]{Yanagida:1979as}%
  \BibitemOpen
  \bibfield  {author} {\bibinfo {author} {\bibfnamefont {T.}~\bibnamefont
  {Yanagida}},\ }in\ \href@noop {} {\emph {\bibinfo {booktitle} {Proc. Workshop
  on the Baryon Number of the Universe and Unified Theories}}},\ \bibinfo
  {editor} {edited by\ \bibinfo {editor} {\bibfnamefont {O.}~\bibnamefont
  {Sawada}}\ and\ \bibinfo {editor} {\bibfnamefont {A.}~\bibnamefont
  {Sugamoto}}}\ (\bibinfo {year} {1979})\ p.~\bibinfo {pages} {95}\BibitemShut
  {NoStop}%
\bibitem [{\citenamefont {Mohapatra}\ and\ \citenamefont
  {Senjanovic}(1980)}]{Mohapatra:1979ia}%
  \BibitemOpen
  \bibfield  {author} {\bibinfo {author} {\bibfnamefont {R.~N.}\ \bibnamefont
  {Mohapatra}}\ and\ \bibinfo {author} {\bibfnamefont {G.}~\bibnamefont
  {Senjanovic}},\ }\bibfield  {journal} {\bibinfo  {journal} {Phys.Rev.Lett.}\
  }\textbf {\bibinfo {volume} {44}},\ \bibinfo {pages} {912} (\bibinfo {year}
  {1980})\BibitemShut {NoStop}%
\bibitem [{\citenamefont {Gell-Mann}\ \emph {et~al.}(1979)\citenamefont
  {Gell-Mann}, \citenamefont {Ramond},\ and\ \citenamefont
  {Slansky}}]{GellMann:1980vs}%
  \BibitemOpen
  \bibfield  {author} {\bibinfo {author} {\bibfnamefont {M.}~\bibnamefont
  {Gell-Mann}}, \bibinfo {author} {\bibfnamefont {P.}~\bibnamefont {Ramond}}, \
  and\ \bibinfo {author} {\bibfnamefont {R.}~\bibnamefont {Slansky}},\ }in\
  \href@noop {} {\emph {\bibinfo {booktitle} {Supergravity}}},\ \bibinfo
  {editor} {edited by\ \bibinfo {editor} {\bibfnamefont {P.}~\bibnamefont {{van
  Nieuwenhuizen}}}\ and\ \bibinfo {editor} {\bibfnamefont {D.}~\bibnamefont
  {Freedman}}}\ (\bibinfo {year} {1979})\ p.\ \bibinfo {pages}
  {315}\BibitemShut {NoStop}%
\bibitem [{\citenamefont {Barry}\ \emph {et~al.}(2011)\citenamefont {Barry},
  \citenamefont {Rodejohann},\ and\ \citenamefont {Zhang}}]{Barry:2011wb}%
  \BibitemOpen
  \bibfield  {author} {\bibinfo {author} {\bibfnamefont {J.}~\bibnamefont
  {Barry}}, \bibinfo {author} {\bibfnamefont {W.}~\bibnamefont {Rodejohann}}, \
  and\ \bibinfo {author} {\bibfnamefont {H.}~\bibnamefont {Zhang}},\ }\bibfield
   {journal} {\bibinfo  {journal} {JHEP}\ }\textbf {\bibinfo {volume} {1107}},\
  \bibinfo {pages} {091} (\bibinfo {year} {2011}),\ \Eprint
  {http://arxiv.org/abs/1105.3911} {1105.3911}\BibitemShut {NoStop}%
\bibitem [{\citenamefont {Zhang}(2012)}]{Zhang:2011vh}%
  \BibitemOpen
  \bibfield  {author} {\bibinfo {author} {\bibfnamefont {H.}~\bibnamefont
  {Zhang}},\ }\bibfield  {journal} {\bibinfo  {journal} {Phys.Lett.}\ }\textbf
  {\bibinfo {volume} {B714}},\ \bibinfo {pages} {262} (\bibinfo {year}
  {2012}),\ \Eprint {http://arxiv.org/abs/1110.6838} {1110.6838}\BibitemShut
  {NoStop}%
\bibitem [{\citenamefont {Ma}\ and\ \citenamefont {Roy}(1995)}]{Ma:1995gf}%
  \BibitemOpen
  \bibfield  {author} {\bibinfo {author} {\bibfnamefont {E.}~\bibnamefont
  {Ma}}\ and\ \bibinfo {author} {\bibfnamefont {P.}~\bibnamefont {Roy}},\
  }\bibfield  {journal} {\bibinfo  {journal} {Phys.Rev.}\ }\textbf {\bibinfo
  {volume} {D52}},\ \bibinfo {pages} {4780} (\bibinfo {year} {1995}),\ \Eprint
  {http://arxiv.org/abs/hep-ph/9504342} {hep-ph/9504342}\BibitemShut {NoStop}%
\bibitem [{\citenamefont {Chun}\ \emph {et~al.}(1995)\citenamefont {Chun},
  \citenamefont {Joshipura},\ and\ \citenamefont {Smirnov}}]{Chun:1995js}%
  \BibitemOpen
  \bibfield  {author} {\bibinfo {author} {\bibfnamefont {E.}~\bibnamefont
  {Chun}}, \bibinfo {author} {\bibfnamefont {A.~S.}\ \bibnamefont {Joshipura}},
  \ and\ \bibinfo {author} {\bibfnamefont {A.~Y.}\ \bibnamefont {Smirnov}},\
  }\bibfield  {journal} {\bibinfo  {journal} {Phys.Lett.}\ }\textbf {\bibinfo
  {volume} {B357}},\ \bibinfo {pages} {608} (\bibinfo {year} {1995}),\ \Eprint
  {http://arxiv.org/abs/hep-ph/9505275} {hep-ph/9505275}\BibitemShut {NoStop}%
\bibitem [{\citenamefont {Schechter}\ and\ \citenamefont
  {Valle}(1980)}]{Schechter:1980gr}%
  \BibitemOpen
  \bibfield  {author} {\bibinfo {author} {\bibfnamefont {J.}~\bibnamefont
  {Schechter}}\ and\ \bibinfo {author} {\bibfnamefont {J.}~\bibnamefont
  {Valle}},\ }\bibfield  {journal} {\bibinfo  {journal} {Phys.Rev.}\ }\textbf
  {\bibinfo {volume} {D22}},\ \bibinfo {pages} {2227} (\bibinfo {year}
  {1980})\BibitemShut {NoStop}%
\bibitem [{\citenamefont {Green}\ and\ \citenamefont
  {Schwarz}(1984)}]{Green:1984sg}%
  \BibitemOpen
  \bibfield  {author} {\bibinfo {author} {\bibfnamefont {M.~B.}\ \bibnamefont
  {Green}}\ and\ \bibinfo {author} {\bibfnamefont {J.~H.}\ \bibnamefont
  {Schwarz}},\ }\bibfield  {journal} {\bibinfo  {journal} {Phys.Lett.}\
  }\textbf {\bibinfo {volume} {B149}},\ \bibinfo {pages} {117} (\bibinfo {year}
  {1984})\BibitemShut {NoStop}%
\bibitem [{\citenamefont {Babu}\ and\ \citenamefont
  {Seidl}(2004{\natexlab{a}})}]{Babu:2003is}%
  \BibitemOpen
  \bibfield  {author} {\bibinfo {author} {\bibfnamefont {K.}~\bibnamefont
  {Babu}}\ and\ \bibinfo {author} {\bibfnamefont {G.}~\bibnamefont {Seidl}},\
  }\bibfield  {journal} {\bibinfo  {journal} {Phys.Lett.}\ }\textbf {\bibinfo
  {volume} {B591}},\ \bibinfo {pages} {127} (\bibinfo {year}
  {2004}{\natexlab{a}}),\ \Eprint {http://arxiv.org/abs/hep-ph/0312285}
  {hep-ph/0312285}\BibitemShut {NoStop}%
\bibitem [{\citenamefont {Babu}\ and\ \citenamefont
  {Seidl}(2004{\natexlab{b}})}]{Babu:2004mj}%
  \BibitemOpen
  \bibfield  {author} {\bibinfo {author} {\bibfnamefont {K.}~\bibnamefont
  {Babu}}\ and\ \bibinfo {author} {\bibfnamefont {G.}~\bibnamefont {Seidl}},\
  }\bibfield  {journal} {\bibinfo  {journal} {Phys.Rev.}\ }\textbf {\bibinfo
  {volume} {D70}},\ \bibinfo {pages} {113014} (\bibinfo {year}
  {2004}{\natexlab{b}}),\ \Eprint {http://arxiv.org/abs/hep-ph/0405197}
  {hep-ph/0405197}\BibitemShut {NoStop}%
\bibitem [{\citenamefont {Sayre}\ \emph {et~al.}(2005)\citenamefont {Sayre},
  \citenamefont {Wiesenfeldt},\ and\ \citenamefont
  {Willenbrock}}]{Sayre:2005yh}%
  \BibitemOpen
  \bibfield  {author} {\bibinfo {author} {\bibfnamefont {J.}~\bibnamefont
  {Sayre}}, \bibinfo {author} {\bibfnamefont {S.}~\bibnamefont {Wiesenfeldt}},
  \ and\ \bibinfo {author} {\bibfnamefont {S.}~\bibnamefont {Willenbrock}},\
  }\bibfield  {journal} {\bibinfo  {journal} {Phys.Rev.}\ }\textbf {\bibinfo
  {volume} {D72}},\ \bibinfo {pages} {015001} (\bibinfo {year} {2005}),\
  \Eprint {http://arxiv.org/abs/hep-ph/0504198} {hep-ph/0504198}\BibitemShut
  {NoStop}%
\bibitem [{\citenamefont {Batell}(2011)}]{Batell:2010bp}%
  \BibitemOpen
  \bibfield  {author} {\bibinfo {author} {\bibfnamefont {B.}~\bibnamefont
  {Batell}},\ }\bibfield  {journal} {\bibinfo  {journal} {Phys.Rev.}\ }\textbf
  {\bibinfo {volume} {D83}},\ \bibinfo {pages} {035006} (\bibinfo {year}
  {2011}),\ \Eprint {http://arxiv.org/abs/1007.0045} {1007.0045}\BibitemShut
  {NoStop}%
\bibitem [{\citenamefont {Batra}\ \emph {et~al.}(2006)\citenamefont {Batra},
  \citenamefont {Dobrescu},\ and\ \citenamefont {Spivak}}]{Batra:2005rh}%
  \BibitemOpen
  \bibfield  {author} {\bibinfo {author} {\bibfnamefont {P.}~\bibnamefont
  {Batra}}, \bibinfo {author} {\bibfnamefont {B.~A.}\ \bibnamefont {Dobrescu}},
  \ and\ \bibinfo {author} {\bibfnamefont {D.}~\bibnamefont {Spivak}},\
  }\bibfield  {journal} {\bibinfo  {journal} {J.Math.Phys.}\ }\textbf {\bibinfo
  {volume} {47}},\ \bibinfo {pages} {082301} (\bibinfo {year} {2006}),\ \Eprint
  {http://arxiv.org/abs/hep-ph/0510181} {hep-ph/0510181}\BibitemShut {NoStop}%
\bibitem [{\citenamefont {Nakayama}\ \emph {et~al.}(2011)\citenamefont
  {Nakayama}, \citenamefont {Takahashi},\ and\ \citenamefont
  {Yanagida}}]{Nakayama:2011dj}%
  \BibitemOpen
  \bibfield  {author} {\bibinfo {author} {\bibfnamefont {K.}~\bibnamefont
  {Nakayama}}, \bibinfo {author} {\bibfnamefont {F.}~\bibnamefont {Takahashi}},
  \ and\ \bibinfo {author} {\bibfnamefont {T.~T.}\ \bibnamefont {Yanagida}},\
  }\bibfield  {journal} {\bibinfo  {journal} {Phys.Lett.}\ }\textbf {\bibinfo
  {volume} {B699}},\ \bibinfo {pages} {360} (\bibinfo {year} {2011}),\ \Eprint
  {http://arxiv.org/abs/1102.4688} {1102.4688}\BibitemShut {NoStop}%
\bibitem [{\citenamefont {Fukugita}\ and\ \citenamefont
  {Yanagida}(1986)}]{Fukugita:1986hr}%
  \BibitemOpen
  \bibfield  {author} {\bibinfo {author} {\bibfnamefont {M.}~\bibnamefont
  {Fukugita}}\ and\ \bibinfo {author} {\bibfnamefont {T.}~\bibnamefont
  {Yanagida}},\ }\bibfield  {journal} {\bibinfo  {journal} {Phys.Lett.}\
  }\textbf {\bibinfo {volume} {B174}},\ \bibinfo {pages} {45} (\bibinfo {year}
  {1986})\BibitemShut {NoStop}%
\bibitem [{\citenamefont {Patt}\ and\ \citenamefont
  {Wilczek}(2006)}]{Patt:2006fw}%
  \BibitemOpen
  \bibfield  {author} {\bibinfo {author} {\bibfnamefont {B.}~\bibnamefont
  {Patt}}\ and\ \bibinfo {author} {\bibfnamefont {F.}~\bibnamefont {Wilczek}}\
  } (\bibinfo {year} {2006}),\ \Eprint {http://arxiv.org/abs/hep-ph/0605188}
  {hep-ph/0605188}\BibitemShut {NoStop}%
\bibitem [{\citenamefont {Holdom}(1986)}]{Holdom:1985ag}%
  \BibitemOpen
  \bibfield  {author} {\bibinfo {author} {\bibfnamefont {B.}~\bibnamefont
  {Holdom}},\ }\bibfield  {journal} {\bibinfo  {journal} {Phys.Lett.}\ }\textbf
  {\bibinfo {volume} {B166}},\ \bibinfo {pages} {196} (\bibinfo {year}
  {1986})\BibitemShut {NoStop}%
\bibitem [{\citenamefont {Holdom}(1991)}]{Holdom:1990xp}%
  \BibitemOpen
  \bibfield  {author} {\bibinfo {author} {\bibfnamefont {B.}~\bibnamefont
  {Holdom}},\ }\bibfield  {journal} {\bibinfo  {journal} {Phys.Lett.}\ }\textbf
  {\bibinfo {volume} {B259}},\ \bibinfo {pages} {329} (\bibinfo {year}
  {1991})\BibitemShut {NoStop}%
\bibitem [{\citenamefont {Belanger}\ \emph {et~al.}(2008)\citenamefont
  {Belanger}, \citenamefont {Pukhov},\ and\ \citenamefont
  {Servant}}]{Belanger:2007dx}%
  \BibitemOpen
  \bibfield  {author} {\bibinfo {author} {\bibfnamefont {G.}~\bibnamefont
  {Belanger}}, \bibinfo {author} {\bibfnamefont {A.}~\bibnamefont {Pukhov}}, \
  and\ \bibinfo {author} {\bibfnamefont {G.}~\bibnamefont {Servant}},\
  }\bibfield  {journal} {\bibinfo  {journal} {JCAP}\ }\textbf {\bibinfo
  {volume} {0801}},\ \bibinfo {pages} {009} (\bibinfo {year} {2008}),\ \Eprint
  {http://arxiv.org/abs/0706.0526} {0706.0526}\BibitemShut {NoStop}%
\bibitem [{\citenamefont {Mambrini}(2011)}]{Mambrini:2011dw}%
  \BibitemOpen
  \bibfield  {author} {\bibinfo {author} {\bibfnamefont {Y.}~\bibnamefont
  {Mambrini}},\ }\bibfield  {journal} {\bibinfo  {journal} {JCAP}\ }\textbf
  {\bibinfo {volume} {1107}},\ \bibinfo {pages} {009} (\bibinfo {year}
  {2011}),\ \Eprint {http://arxiv.org/abs/1104.4799} {1104.4799}\BibitemShut
  {NoStop}%
\bibitem [{\citenamefont {Chu}\ \emph {et~al.}(2012)\citenamefont {Chu},
  \citenamefont {Hambye},\ and\ \citenamefont {Tytgat}}]{Chu:2011be}%
  \BibitemOpen
  \bibfield  {author} {\bibinfo {author} {\bibfnamefont {X.}~\bibnamefont
  {Chu}}, \bibinfo {author} {\bibfnamefont {T.}~\bibnamefont {Hambye}}, \ and\
  \bibinfo {author} {\bibfnamefont {M.~H.}\ \bibnamefont {Tytgat}},\ }\bibfield
   {journal} {\bibinfo  {journal} {JCAP}\ }\textbf {\bibinfo {volume} {1205}},\
  \bibinfo {pages} {034} (\bibinfo {year} {2012}),\ \Eprint
  {http://arxiv.org/abs/1112.0493} {1112.0493}\BibitemShut {NoStop}%
\bibitem [{\citenamefont {Falkowski}\ \emph {et~al.}(2009)\citenamefont
  {Falkowski}, \citenamefont {Juknevich},\ and\ \citenamefont
  {Shelton}}]{Falkowski:2009yz}%
  \BibitemOpen
  \bibfield  {author} {\bibinfo {author} {\bibfnamefont {A.}~\bibnamefont
  {Falkowski}}, \bibinfo {author} {\bibfnamefont {J.}~\bibnamefont
  {Juknevich}}, \ and\ \bibinfo {author} {\bibfnamefont {J.}~\bibnamefont
  {Shelton}}\ } (\bibinfo {year} {2009}),\ \Eprint
  {http://arxiv.org/abs/0908.1790} {0908.1790}\BibitemShut {NoStop}%
\bibitem [{\citenamefont {Lindner}\ \emph {et~al.}(2010)\citenamefont
  {Lindner}, \citenamefont {Merle},\ and\ \citenamefont
  {Niro}}]{Lindner:2010rr}%
  \BibitemOpen
  \bibfield  {author} {\bibinfo {author} {\bibfnamefont {M.}~\bibnamefont
  {Lindner}}, \bibinfo {author} {\bibfnamefont {A.}~\bibnamefont {Merle}}, \
  and\ \bibinfo {author} {\bibfnamefont {V.}~\bibnamefont {Niro}},\ }\bibfield
  {journal} {\bibinfo  {journal} {Phys.Rev.}\ }\textbf {\bibinfo {volume}
  {D82}},\ \bibinfo {pages} {123529} (\bibinfo {year} {2010}),\ \Eprint
  {http://arxiv.org/abs/1005.3116} {1005.3116}\BibitemShut {NoStop}%
\bibitem [{\citenamefont {Farzan}(2012)}]{Farzan:2011ck}%
  \BibitemOpen
  \bibfield  {author} {\bibinfo {author} {\bibfnamefont {Y.}~\bibnamefont
  {Farzan}},\ }\bibfield  {journal} {\bibinfo  {journal} {JHEP}\ }\textbf
  {\bibinfo {volume} {1202}},\ \bibinfo {pages} {091} (\bibinfo {year}
  {2012}),\ \Eprint {http://arxiv.org/abs/1111.1063} {1111.1063}\BibitemShut
  {NoStop}%
\bibitem [{\citenamefont {Boehm}\ \emph {et~al.}(2004)\citenamefont {Boehm},
  \citenamefont {Fayet},\ and\ \citenamefont {Silk}}]{Boehm:2003ha}%
  \BibitemOpen
  \bibfield  {author} {\bibinfo {author} {\bibfnamefont {C.}~\bibnamefont
  {Boehm}}, \bibinfo {author} {\bibfnamefont {P.}~\bibnamefont {Fayet}}, \ and\
  \bibinfo {author} {\bibfnamefont {J.}~\bibnamefont {Silk}},\ }\bibfield
  {journal} {\bibinfo  {journal} {Phys.Rev.}\ }\textbf {\bibinfo {volume}
  {D69}},\ \bibinfo {pages} {101302} (\bibinfo {year} {2004}),\ \Eprint
  {http://arxiv.org/abs/hep-ph/0311143} {hep-ph/0311143}\BibitemShut {NoStop}%
\bibitem [{\citenamefont {Ma}(2006)}]{Ma:2006uv}%
  \BibitemOpen
  \bibfield  {author} {\bibinfo {author} {\bibfnamefont {E.}~\bibnamefont
  {Ma}},\ }\bibfield  {journal} {\bibinfo  {journal} {Annales Fond.Broglie}\
  }\textbf {\bibinfo {volume} {31}},\ \bibinfo {pages} {285} (\bibinfo {year}
  {2006}),\ \Eprint {http://arxiv.org/abs/hep-ph/0607142}
  {hep-ph/0607142}\BibitemShut {NoStop}%
\bibitem [{\citenamefont {Hur}\ \emph {et~al.}(2008)\citenamefont {Hur},
  \citenamefont {Lee},\ and\ \citenamefont {Nasri}}]{Hur:2007ur}%
  \BibitemOpen
  \bibfield  {author} {\bibinfo {author} {\bibfnamefont {T.}~\bibnamefont
  {Hur}}, \bibinfo {author} {\bibfnamefont {H.-S.}\ \bibnamefont {Lee}}, \ and\
  \bibinfo {author} {\bibfnamefont {S.}~\bibnamefont {Nasri}},\ }\bibfield
  {journal} {\bibinfo  {journal} {Phys.Rev.}\ }\textbf {\bibinfo {volume}
  {D77}},\ \bibinfo {pages} {015008} (\bibinfo {year} {2008}),\ \Eprint
  {http://arxiv.org/abs/0710.2653} {0710.2653}\BibitemShut {NoStop}%
\bibitem [{\citenamefont {Cao}\ \emph {et~al.}(2007)\citenamefont {Cao},
  \citenamefont {Ma}, \citenamefont {Wudka},\ and\ \citenamefont
  {Yuan}}]{Cao:2007fy}%
  \BibitemOpen
  \bibfield  {author} {\bibinfo {author} {\bibfnamefont {Q.-H.}\ \bibnamefont
  {Cao}}, \bibinfo {author} {\bibfnamefont {E.}~\bibnamefont {Ma}}, \bibinfo
  {author} {\bibfnamefont {J.}~\bibnamefont {Wudka}}, \ and\ \bibinfo {author}
  {\bibfnamefont {C.-P.}\ \bibnamefont {Yuan}}\ } (\bibinfo {year} {2007}),\
  \Eprint {http://arxiv.org/abs/0711.3881} {0711.3881}\BibitemShut {NoStop}%
\bibitem [{\citenamefont {Weinberg}(1979)}]{Weinberg:1979sa}%
  \BibitemOpen
  \bibfield  {author} {\bibinfo {author} {\bibfnamefont {S.}~\bibnamefont
  {Weinberg}},\ }\bibfield  {journal} {\bibinfo  {journal} {Phys.Rev.Lett.}\
  }\textbf {\bibinfo {volume} {43}},\ \bibinfo {pages} {1566} (\bibinfo {year}
  {1979})\BibitemShut {NoStop}%
\bibitem [{\citenamefont {Lopez-Honorez}\ \emph {et~al.}(2012)\citenamefont
  {Lopez-Honorez}, \citenamefont {Schwetz},\ and\ \citenamefont
  {Zupan}}]{LopezHonorez:2012kv}%
  \BibitemOpen
  \bibfield  {author} {\bibinfo {author} {\bibfnamefont {L.}~\bibnamefont
  {Lopez-Honorez}}, \bibinfo {author} {\bibfnamefont {T.}~\bibnamefont
  {Schwetz}}, \ and\ \bibinfo {author} {\bibfnamefont {J.}~\bibnamefont
  {Zupan}},\ }\bibfield  {journal} {\bibinfo  {journal} {Phys.Lett.}\ }\textbf
  {\bibinfo {volume} {B716}},\ \bibinfo {pages} {179} (\bibinfo {year}
  {2012}),\ \Eprint {http://arxiv.org/abs/1203.2064} {1203.2064}\BibitemShut
  {NoStop}%
\bibitem [{\citenamefont {Heeck}\ and\ \citenamefont {Zhang}()}]{preparation}%
  \BibitemOpen
  \bibfield  {author} {\bibinfo {author} {\bibfnamefont {J.}~\bibnamefont
  {Heeck}}\ and\ \bibinfo {author} {\bibfnamefont {H.}~\bibnamefont {Zhang}},\
  }\href@noop {} {\emph { \bibinfo {title} {{in preparation}}}}\BibitemShut {NoStop}%
\bibitem [{\citenamefont {Joudaki}\ \emph {et~al.}(2013)\citenamefont
  {Joudaki}, \citenamefont {Abazajian},\ and\ \citenamefont
  {Kaplinghat}}]{Joudaki:2012uk}%
  \BibitemOpen
  \bibfield  {author} {\bibinfo {author} {\bibfnamefont {S.}~\bibnamefont
  {Joudaki}}, \bibinfo {author} {\bibfnamefont {K.~N.}\ \bibnamefont
  {Abazajian}}, \ and\ \bibinfo {author} {\bibfnamefont {M.}~\bibnamefont
  {Kaplinghat}},\ }\bibfield  {journal} {\bibinfo  {journal} {Phys. Rev.}\
  }\textbf {\bibinfo {volume} {D87}},\ \bibinfo {pages} {065003} (\bibinfo
  {year} {2013}),\ \Eprint {http://arxiv.org/abs/1208.4354}
  {1208.4354}\BibitemShut {NoStop}%
\bibitem [{\citenamefont {Motohashi}\ \emph {et~al.}(2013)\citenamefont
  {Motohashi}, \citenamefont {Starobinsky},\ and\ \citenamefont
  {Yokoyama}}]{Motohashi:2012wc}%
  \BibitemOpen
  \bibfield  {author} {\bibinfo {author} {\bibfnamefont {H.}~\bibnamefont
  {Motohashi}}, \bibinfo {author} {\bibfnamefont {A.~A.}\ \bibnamefont
  {Starobinsky}}, \ and\ \bibinfo {author} {\bibfnamefont {J.}~\bibnamefont
  {Yokoyama}},\ }\bibfield  {journal} {\bibinfo  {journal} {Phys. Rev. Lett.}\
  }\textbf {\bibinfo {volume} {110}},\ \bibinfo {pages} {121302} (\bibinfo
  {year} {2013}),\ \Eprint {http://arxiv.org/abs/1203.6828}
  {1203.6828}\BibitemShut {NoStop}%
\bibitem [{\citenamefont {Abazajian}\ \emph {et~al.}(2005)\citenamefont
  {Abazajian}, \citenamefont {Bell}, \citenamefont {Fuller},\ and\
  \citenamefont {Wong}}]{Abazajian:2004aj}%
  \BibitemOpen
  \bibfield  {author} {\bibinfo {author} {\bibfnamefont {K.}~\bibnamefont
  {Abazajian}}, \bibinfo {author} {\bibfnamefont {N.~F.}\ \bibnamefont {Bell}},
  \bibinfo {author} {\bibfnamefont {G.~M.}\ \bibnamefont {Fuller}}, \ and\
  \bibinfo {author} {\bibfnamefont {Y.~Y.}\ \bibnamefont {Wong}},\ }\bibfield
  {journal} {\bibinfo  {journal} {Phys.Rev.}\ }\textbf {\bibinfo {volume}
  {D72}},\ \bibinfo {pages} {063004} (\bibinfo {year} {2005}),\ \Eprint
  {http://arxiv.org/abs/astro-ph/0410175} {astro-ph/0410175}\BibitemShut
  {NoStop}%
\bibitem [{\citenamefont {Melchiorri}\ \emph {et~al.}(2009)\citenamefont
  {Melchiorri}, \citenamefont {Mena}, \citenamefont {Palomares-Ruiz},
  \citenamefont {Pascoli}, \citenamefont {Slosar} \emph
  {et~al.}}]{Melchiorri:2008gq}%
  \BibitemOpen
  \bibfield  {author} {\bibinfo {author} {\bibfnamefont {A.}~\bibnamefont
  {Melchiorri}}, \bibinfo {author} {\bibfnamefont {O.}~\bibnamefont {Mena}},
  \bibinfo {author} {\bibfnamefont {S.}~\bibnamefont {Palomares-Ruiz}},
  \bibinfo {author} {\bibfnamefont {S.}~\bibnamefont {Pascoli}}, \bibinfo
  {author} {\bibfnamefont {A.}~\bibnamefont {Slosar}},  \emph {et~al.},\
  }\bibfield  {journal} {\bibinfo  {journal} {JCAP}\ }\textbf {\bibinfo
  {volume} {0901}},\ \bibinfo {pages} {036} (\bibinfo {year} {2009}),\ \Eprint
  {http://arxiv.org/abs/0810.5133} {0810.5133}\BibitemShut {NoStop}%
\bibitem [{\citenamefont {Giunti}\ and\ \citenamefont
  {Laveder}(2011{\natexlab{b}})}]{Giunti:2011cp}%
  \BibitemOpen
  \bibfield  {author} {\bibinfo {author} {\bibfnamefont {C.}~\bibnamefont
  {Giunti}}\ and\ \bibinfo {author} {\bibfnamefont {M.}~\bibnamefont
  {Laveder}},\ }\bibfield  {journal} {\bibinfo  {journal} {Phys.Lett.}\
  }\textbf {\bibinfo {volume} {B706}},\ \bibinfo {pages} {200} (\bibinfo {year}
  {2011}{\natexlab{b}}),\ \Eprint {http://arxiv.org/abs/1111.1069}
  {1111.1069}\BibitemShut {NoStop}%
\bibitem [{\citenamefont {Hannestad}\ \emph {et~al.}(2012)\citenamefont
  {Hannestad}, \citenamefont {Tamborra},\ and\ \citenamefont
  {Tram}}]{Hannestad:2012ky}%
  \BibitemOpen
  \bibfield  {author} {\bibinfo {author} {\bibfnamefont {S.}~\bibnamefont
  {Hannestad}}, \bibinfo {author} {\bibfnamefont {I.}~\bibnamefont {Tamborra}},
  \ and\ \bibinfo {author} {\bibfnamefont {T.}~\bibnamefont {Tram}},\
  }\bibfield  {journal} {\bibinfo  {journal} {JCAP}\ }\textbf {\bibinfo
  {volume} {1207}},\ \bibinfo {pages} {025} (\bibinfo {year} {2012}),\ \Eprint
  {http://arxiv.org/abs/1204.5861} {1204.5861}\BibitemShut {NoStop}%
\bibitem [{\citenamefont {Archidiacono}\ \emph {et~al.}(2012)\citenamefont
  {Archidiacono}, \citenamefont {Fornengo}, \citenamefont {Giunti},\ and\
  \citenamefont {Melchiorri}}]{Archidiacono:2012ri}%
  \BibitemOpen
  \bibfield  {author} {\bibinfo {author} {\bibfnamefont {M.}~\bibnamefont
  {Archidiacono}}, \bibinfo {author} {\bibfnamefont {N.}~\bibnamefont
  {Fornengo}}, \bibinfo {author} {\bibfnamefont {C.}~\bibnamefont {Giunti}}, \
  and\ \bibinfo {author} {\bibfnamefont {A.}~\bibnamefont {Melchiorri}},\
  }\bibfield  {journal} {\bibinfo  {journal} {Phys.Rev.}\ }\textbf {\bibinfo
  {volume} {D86}},\ \bibinfo {pages} {065028} (\bibinfo {year} {2012}),\
  \Eprint {http://arxiv.org/abs/1207.6515} {1207.6515}\BibitemShut {NoStop}%
\bibitem [{\citenamefont {Komatsu}\ \emph {et~al.}(2011)\citenamefont {Komatsu}
  \emph {et~al.}}]{Komatsu:2010fb}%
  \BibitemOpen
  \bibfield  {author} {\bibinfo {author} {\bibfnamefont {E.}~\bibnamefont
  {Komatsu}} \emph {et~al.} (\bibinfo {collaboration} {WMAP Collaboration}),\
  }\bibfield  {journal} {\bibinfo  {journal} {Astrophys.J.Suppl.}\ }\textbf
  {\bibinfo {volume} {192}},\ \bibinfo {pages} {18} (\bibinfo {year} {2011}),\
  \Eprint {http://arxiv.org/abs/1001.4538} {1001.4538}\BibitemShut {NoStop}%
\bibitem [{\citenamefont {Larson}\ \emph {et~al.}(2011)\citenamefont {Larson},
  \citenamefont {Dunkley}, \citenamefont {Hinshaw}, \citenamefont {Komatsu},
  \citenamefont {Nolta} \emph {et~al.}}]{Larson:2010gs}%
  \BibitemOpen
  \bibfield  {author} {\bibinfo {author} {\bibfnamefont {D.}~\bibnamefont
  {Larson}}, \bibinfo {author} {\bibfnamefont {J.}~\bibnamefont {Dunkley}},
  \bibinfo {author} {\bibfnamefont {G.}~\bibnamefont {Hinshaw}}, \bibinfo
  {author} {\bibfnamefont {E.}~\bibnamefont {Komatsu}}, \bibinfo {author}
  {\bibfnamefont {M.}~\bibnamefont {Nolta}},  \emph {et~al.},\ }\bibfield
  {journal} {\bibinfo  {journal} {Astrophys.J.Suppl.}\ }\textbf {\bibinfo
  {volume} {192}},\ \bibinfo {pages} {16} (\bibinfo {year} {2011}),\ \Eprint
  {http://arxiv.org/abs/1001.4635} {1001.4635}\BibitemShut {NoStop}%
\bibitem [{\citenamefont {Belanger}\ \emph {et~al.}(2007)\citenamefont
  {Belanger}, \citenamefont {Boudjema}, \citenamefont {Pukhov},\ and\
  \citenamefont {Semenov}}]{Belanger:2006is}%
  \BibitemOpen
  \bibfield  {author} {\bibinfo {author} {\bibfnamefont {G.}~\bibnamefont
  {Belanger}}, \bibinfo {author} {\bibfnamefont {F.}~\bibnamefont {Boudjema}},
  \bibinfo {author} {\bibfnamefont {A.}~\bibnamefont {Pukhov}}, \ and\ \bibinfo
  {author} {\bibfnamefont {A.}~\bibnamefont {Semenov}},\ }\bibfield  {journal}
  {\bibinfo  {journal} {Comput.Phys.Commun.}\ }\textbf {\bibinfo {volume}
  {176}},\ \bibinfo {pages} {367} (\bibinfo {year} {2007}),\ \Eprint
  {http://arxiv.org/abs/hep-ph/0607059} {hep-ph/0607059}\BibitemShut {NoStop}%
\bibitem [{\citenamefont {Belanger}\ \emph {et~al.}(2009)\citenamefont
  {Belanger}, \citenamefont {Boudjema}, \citenamefont {Pukhov},\ and\
  \citenamefont {Semenov}}]{Belanger:2008sj}%
  \BibitemOpen
  \bibfield  {author} {\bibinfo {author} {\bibfnamefont {G.}~\bibnamefont
  {Belanger}}, \bibinfo {author} {\bibfnamefont {F.}~\bibnamefont {Boudjema}},
  \bibinfo {author} {\bibfnamefont {A.}~\bibnamefont {Pukhov}}, \ and\ \bibinfo
  {author} {\bibfnamefont {A.}~\bibnamefont {Semenov}},\ }\bibfield  {journal}
  {\bibinfo  {journal} {Comput.Phys.Commun.}\ }\textbf {\bibinfo {volume}
  {180}},\ \bibinfo {pages} {747} (\bibinfo {year} {2009}),\ \Eprint
  {http://arxiv.org/abs/0803.2360} {0803.2360}\BibitemShut {NoStop}%
\bibitem [{\citenamefont {Belanger}\ \emph {et~al.}(2010)\citenamefont
  {Belanger}, \citenamefont {Boudjema}, \citenamefont {Pukhov},\ and\
  \citenamefont {Semenov}}]{Belanger:2010pz}%
  \BibitemOpen
  \bibfield  {author} {\bibinfo {author} {\bibfnamefont {G.}~\bibnamefont
  {Belanger}}, \bibinfo {author} {\bibfnamefont {F.}~\bibnamefont {Boudjema}},
  \bibinfo {author} {\bibfnamefont {A.}~\bibnamefont {Pukhov}}, \ and\ \bibinfo
  {author} {\bibfnamefont {A.}~\bibnamefont {Semenov}},\ }\bibfield  {journal}
  {\bibinfo  {journal} {Nuovo Cim.}\ }\textbf {\bibinfo {volume} {C033N2}},\
  \bibinfo {pages} {111} (\bibinfo {year} {2010}),\ \Eprint
  {http://arxiv.org/abs/1005.4133} {1005.4133}\BibitemShut {NoStop}%
\bibitem [{\citenamefont {Abbasi}\ \emph {et~al.}(2011)\citenamefont {Abbasi}
  \emph {et~al.}}]{Abbasi:2011eq}%
  \BibitemOpen
  \bibfield  {author} {\bibinfo {author} {\bibfnamefont {R.}~\bibnamefont
  {Abbasi}} \emph {et~al.} (\bibinfo {collaboration} {IceCube Collaboration}),\
  }\bibfield  {journal} {\bibinfo  {journal} {Phys.Rev.}\ }\textbf {\bibinfo
  {volume} {D84}},\ \bibinfo {pages} {022004} (\bibinfo {year} {2011}),\
  \Eprint {http://arxiv.org/abs/1101.3349} {1101.3349}\BibitemShut {NoStop}%
\bibitem [{\citenamefont {Abbasi}\ \emph {et~al.}(2012)\citenamefont {Abbasi}
  \emph {et~al.}}]{Abbasi:2012ws}%
  \BibitemOpen
  \bibfield  {author} {\bibinfo {author} {\bibfnamefont {R.}~\bibnamefont
  {Abbasi}} \emph {et~al.} (\bibinfo {collaboration} {IceCube collaboration})\
  } (\bibinfo {year} {2012}),\ \Eprint {http://arxiv.org/abs/1210.3557}
  {1210.3557}\BibitemShut {NoStop}%
\bibitem [{\citenamefont {Karagiorgi}\ \emph {et~al.}(2007)\citenamefont
  {Karagiorgi}, \citenamefont {Aguilar-Arevalo}, \citenamefont {Conrad},
  \citenamefont {Shaevitz}, \citenamefont {Whisnant} \emph
  {et~al.}}]{Karagiorgi:2006jf}%
  \BibitemOpen
  \bibfield  {author} {\bibinfo {author} {\bibfnamefont {G.}~\bibnamefont
  {Karagiorgi}}, \bibinfo {author} {\bibfnamefont {A.}~\bibnamefont
  {Aguilar-Arevalo}}, \bibinfo {author} {\bibfnamefont {J.}~\bibnamefont
  {Conrad}}, \bibinfo {author} {\bibfnamefont {M.}~\bibnamefont {Shaevitz}},
  \bibinfo {author} {\bibfnamefont {K.}~\bibnamefont {Whisnant}},  \emph
  {et~al.},\ }\bibfield  {journal} {\bibinfo  {journal} {Phys.Rev.}\ }\textbf
  {\bibinfo {volume} {D75}},\ \bibinfo {pages} {013011} (\bibinfo {year}
  {2007}),\ \Eprint {http://arxiv.org/abs/hep-ph/0609177}
  {hep-ph/0609177}\BibitemShut {NoStop}%
\bibitem [{\citenamefont {Donini}\ \emph {et~al.}(2012)\citenamefont {Donini},
  \citenamefont {Hernandez}, \citenamefont {Lopez-Pavon}, \citenamefont
  {Maltoni},\ and\ \citenamefont {Schwetz}}]{Donini:2012tt}%
  \BibitemOpen
  \bibfield  {author} {\bibinfo {author} {\bibfnamefont {A.}~\bibnamefont
  {Donini}}, \bibinfo {author} {\bibfnamefont {P.}~\bibnamefont {Hernandez}},
  \bibinfo {author} {\bibfnamefont {J.}~\bibnamefont {Lopez-Pavon}}, \bibinfo
  {author} {\bibfnamefont {M.}~\bibnamefont {Maltoni}}, \ and\ \bibinfo
  {author} {\bibfnamefont {T.}~\bibnamefont {Schwetz}},\ }\bibfield  {journal}
  {\bibinfo  {journal} {JHEP}\ }\textbf {\bibinfo {volume} {1207}},\ \bibinfo
  {pages} {161} (\bibinfo {year} {2012}),\ \Eprint
  {http://arxiv.org/abs/1205.5230} {1205.5230}\BibitemShut {NoStop}%
\bibitem [{\citenamefont {Conrad}\ \emph {et~al.}(2013)\citenamefont {Conrad},
  \citenamefont {Ignarra}, \citenamefont {Karagiorgi}, \citenamefont
  {Shaevitz},\ and\ \citenamefont {Spitz}}]{Conrad:2012qt}%
  \BibitemOpen
  \bibfield  {author} {\bibinfo {author} {\bibfnamefont {J.}~\bibnamefont
  {Conrad}}, \bibinfo {author} {\bibfnamefont {C.}~\bibnamefont {Ignarra}},
  \bibinfo {author} {\bibfnamefont {G.}~\bibnamefont {Karagiorgi}}, \bibinfo
  {author} {\bibfnamefont {M.}~\bibnamefont {Shaevitz}}, \ and\ \bibinfo
  {author} {\bibfnamefont {J.}~\bibnamefont {Spitz}},\ }\bibfield  {journal}
  {\bibinfo  {journal} {Adv.High Energy Phys.}\ }\textbf {\bibinfo {volume}
  {2013}},\ \bibinfo {pages} {163897} (\bibinfo {year} {2013}),\ \Eprint
  {http://arxiv.org/abs/1207.4765} {1207.4765}\BibitemShut {NoStop}%
\bibitem [{\citenamefont {Pospelov}(2011)}]{Pospelov:2011ha}%
  \BibitemOpen
  \bibfield  {author} {\bibinfo {author} {\bibfnamefont {M.}~\bibnamefont
  {Pospelov}},\ }\bibfield  {journal} {\bibinfo  {journal} {Phys.Rev.}\
  }\textbf {\bibinfo {volume} {D84}},\ \bibinfo {pages} {085008} (\bibinfo
  {year} {2011}),\ \Eprint {http://arxiv.org/abs/1103.3261}
  {1103.3261}\BibitemShut {NoStop}%
\bibitem [{\citenamefont {Pospelov}\ and\ \citenamefont
  {Pradler}(2012)}]{Pospelov:2012gm}%
  \BibitemOpen
  \bibfield  {author} {\bibinfo {author} {\bibfnamefont {M.}~\bibnamefont
  {Pospelov}}\ and\ \bibinfo {author} {\bibfnamefont {J.}~\bibnamefont
  {Pradler}},\ }\bibfield  {journal} {\bibinfo  {journal} {Phys.Rev.}\ }\textbf
  {\bibinfo {volume} {D85}},\ \bibinfo {pages} {113016} (\bibinfo {year}
  {2012}),\ \Eprint {http://arxiv.org/abs/1203.0545} {1203.0545}\BibitemShut
  {NoStop}%
\end{thebibliography}%
\bibliographystyle{apsrev4-1mod}

\end{document}